\documentclass[journal,a4paper,10pt, romanappendices]{IEEEtran}
\usepackage{balance}
\usepackage{graphicx}
\usepackage{bm}
\usepackage{amsmath}
\usepackage{amsfonts}
\usepackage{amssymb}
\usepackage{color}
\usepackage{mathrsfs}
\usepackage{upgreek}
\usepackage[utf8]{inputenc}
\usepackage{tikz}
\usetikzlibrary{arrows, calc, decorations.markings, positioning}
\usepackage{cite}
\usepackage{booktabs}
\usepackage{multirow}
\usepackage{amsmath}
\usepackage{algorithm}
\usepackage{algorithmic}
\usepackage{makecell}
\usepackage[flushleft]{threeparttable}
\usepackage{midfloat}
\usepackage{tabularx}
\usepackage{flushend}
\usepackage{threeparttable}
 \usepackage[caption=false,font=footnotesize]{subfig}
\usepackage{amssymb}
\usepackage{amsmath}
\usepackage{overpic}
\usepackage{algorithm}
\usepackage{algorithmic}
\usepackage{multirow}
\usepackage{array}
\usepackage{graphicx}
\usepackage{bm}
\usepackage{color}
\usepackage{tikz}
\usepackage{epstopdf}
\usepackage{stfloats}
\usepackage{cite}
\usepackage{tikz}
\usepackage{amsmath}
\usepackage{psfrag}
\usepackage{acronym}
\usepackage{enumerate}
\usetikzlibrary{arrows}
\usetikzlibrary{decorations}

\definecolor{BLUE}{rgb}{0,0,1}

\newtheorem{corollary}{Corollary}
\newtheorem{proposition}{Proposition}
\newtheorem{remark}{Remark}
\newtheorem{lemma}{Lemma}

\newtheorem{assumption}{Assumption}

\acrodef{aoa}[AOA]{angle-of-arrival}
\acrodef{bcrb}[BCRB]{Bayesian Cram\'{e}r-Rao bound}
\acrodef{bfim}[BFIM]{Bayesian Fisher information matrix}
\acrodef{bp}[BP]{belief propagation}
\acrodef{cdi}[CDI]{cooperative dilution intensity}
\acrodef{cir}[CIR]{channel impulse response}
\acrodef{cl}[CL]{cooperative localization}
\acrodef{cp}[CP]{cyclic prefix}
\acrodef{crb}[CRB]{Cram\'{e}r-Rao bound}
\acrodef{crlb}[CRLB]{Cram\'{e}r-Rao lower bound}
\acrodef{dft}[DFT]{discrete Fourier transform}
\acrodef{dof}[DoF]{degree of freedom}
\acrodef{dpeb}[DPEB]{directional position error bound}
\acrodef{fim}[FIM]{Fisher information matrix}
\acrodef{efim}[EFIM]{equivalent Fisher information matrix}
\acrodef{hogs}[HOGS]{hybrid orthogonal-Gaussian signalling}
\acrodef{hsug}[HSUG]{hybrid semi-unitary--Gaussian}
\acrodef{ici}[ICI]{information coupling intensity}
\acrodef{icrb}[ICRB]{inverse CRB}
\acrodef{iid}[i.i.d.]{independently and identically distributed}
\acrodef{im}[IM]{index modulation}
\acrodef{isac}[ISAC]{integrated sensing and communication}
\acrodef{los}[LoS]{line-of-sight}
\acrodef{mse}[MSE]{mean-squared error}
\acrodef{ofdm}[OFDM]{orthogonal frequency-division multiplexing}
\acrodef{pdf}[PDF]{probability density function}
\acrodef{peb}[PEB]{position error bound}
\acrodef{speb}[SPEB]{squared position error bound}
\acrodef{pll}[PLL]{phase-locked loop}
\acrodef{psk}[PSK]{phase shift keying}
\acrodef{p2p}[P2P]{point-to-point}
\acrodef{qam}[QAM]{quadrature amplitude modulation}
\acrodef{rbs}[RBS]{reference broadcast synchronization}
\acrodef{rhs}[RHS]{right hand side}
\acrodef{rii}[RII]{ranging information intensity}
\acrodef{rss}[RSS]{received signal strength}
\acrodef{rc}[RC]{ranging coefficient}
\acrodef{speb}[SPEB]{squared position error bound}
\acrodef{toa}[TOA]{time-of-arrival}
\acrodef{tdoa}[TDOA]{time-difference-of-arrival}
\acrodef{tpsn}[TPSN]{time synchronization protocol for sensor network}
\acrodef{vmp}[VMP]{variational message passing}
\acrodef{wsn}[WSN]{wireless sensor network}
\acrodef{efim}[EFIM]{equivalent Fisher information matrix}
\acrodef{dio}[DIO]{distance-information-only}
\acrodef{aio}[AIO]{angle-information-only}
\acrodef{saaf}[SAAF]{squared array aperture function}
\acrodef{snc}[S\&C]{sensing and communications}
\acrodef{uoa}[UOA]{uniformly oriented array}
\acrodef{rgg}[RGG]{random geometric graph}
\acrodef{snr}[SNR]{signal-to-noise ratio}
\acrodef{eoc}[EoC]{efficiency of cooperation}
\acrodef{npi}[NPI]{nominal position information}
\acrodef{gnss}[GNSS]{global navigation satellite system}
\acrodef{mimo}[MIMO]{multiple-input multiple-output}
\acrodef{mcs}[MCS]{minimally constrained system}
\acrodef{zzb}[ZZB]{Ziv-Zakai lower bound}
\acrodef{wwb}[WWB]{Weiss-Weinstein lower bound}
\acrodef{nlos}[NLOS]{non-light-of-sight}
\acrodef{mmse}[MMSE]{minimum mean squared error}
\acrodef{uav}[UAV]{unmanned aerial vehicle}
\acrodef{ppp}[PPP]{Poisson point process}
\acrodef{bpp}[BPP]{binomial point process}
\acrodef{cln}[CLN]{cooperative location-aware network}
\acrodef{pdr}[PDR]{pedestrian dead reckoning}
\acrodef{ml}[ML]{maximum likelihood}
\acrodef{map}[MAP]{maximum \textit{a posteriori}}
\acrodef{kkt}[KKT]{Karush-Kuhn-Tucker}
\acrodef{st}[ST]{subspace tradeoff}
\acrodef{drt}[DRT]{deterministic-random tradeoff}
\acrodef{ustm}[USTM]{unitary space-time modulation}
\acrodef{pri}[PRI]{pulse repetition interval}
\acrodef{prf}[PRF]{pulse repetition frequency}
\acrodef{acf}[ACF]{autocorrelation function}
\acrodef{isl}[ISL]{integrated sidelobe level}
\acrodef{islr}[ISLR]{integrated sidelobe level ratio}
\acrodef{rgi}[RGI]{range glint intensity}
\acrodef{irgi}[ARGI]{averaged range glint intensity}
\acrodef{cds}[CDS]{cyclic difference set}
\acrodef{psl}[PSL]{peak sidelobe level}
\acrodef{aesl}[AESL]{average expected sidelobe level}
\acrodef{pesl}[PESL]{peak expected sidelobe level}
\acrodef{masm}[MASM]{MASked Modulation}
\acrodef{si}[SI]{self-interference}

\acrodefplural{dof}[DoFs]{degrees of freedom}
\acrodefplural{drt}[DRTs]{deterministic-random tradeoffs}
\acrodefplural{cds}[CDSs]{cyclic difference sets}

\usepackage{hyperref}					% Reference
\usepackage{winsnotation}
\hypersetup{colorlinks,%				% Reference color setup	
	linkcolor=purple,%
	anchorcolor=blue,
    citecolor=purple}

\title{Masked Modulation: High-Throughput Half-Duplex ISAC Transmission Waveform Design}
\author{Yifeng Xiong, \IEEEmembership{Member, IEEE}, Junsheng Mu, \IEEEmembership{Member, IEEE}, Shuangyang Li, \IEEEmembership{Member, IEEE}, \\Marco Lops, \IEEEmembership{Fellow, IEEE}, and Jianhua Zhang, \IEEEmembership{Senior Member, IEEE}
}

\begin{document}
\maketitle

\begin{abstract}
\Ac{isac} enables numerous innovative wireless applications. Communication-centric design is a practical choice for the construction of the sixth generation (6G) \ac{isac} networks. Continuous-wave-based \ac{isac} systems, with \ac{ofdm} being a representative example, suffer from the \ac{si} problem, and hence are less suitable for long-range sensing. On the other hand, pulse-based half-duplex \ac{isac} systems are free of \ac{si}, but are also less favourable for high-throughput communication scenarios. 

In this treatise, we propose \ac{masm}, a half-duplex \ac{isac} waveform design scheme, which minimises a range blindness metric, termed as ``mainlobe fluctuation'', given a duty cycle (proportional to communication throughput) constraint. In particular, \ac{masm} is capable of supporting high-throughput communication ($\sim$50\% duty cycle) under mild mainlobe fluctuation. Moreover, \ac{masm} can be flexibly adapted to frame-level waveform designs by operating on the slow-time scale. In terms of optimal transmit mask design, a set of masks is shown to be \emph{ideal} in the sense of sidelobe level and mainlobe fluctuation intensity. 
\end{abstract}


\section{Introduction}
\IEEEPARstart{I}{ntegrated} sensing and communication (ISAC) has been envisioned by the International Telecommunication Union as one of the six major usage scenarios in the sixth generation (6G) wireless networks \cite{ITU2023}. By coordinating wireless resources such as waveform, hardware platform, time, bandwidth, beam, and energy in a unified manner, \ac{isac} provides an elegant and resource-efficient solution to emerging applications requiring both sensing and communication services, including autonomous driving, low-altitude economy, and human activity sensing \cite{Chafii2023CST,saad2019vision,yuanhaoNW,has_isac}. 

Among all \ac{isac} design paradigms, the communication-centric strategy is arguably the most favourable approach to 6G \ac{isac} networks, which aims for implementing sensing functionalities by imposing minimal modifications on existing communication infrastructures and protocols. Since the publication of the seminal contribution \cite{sturm2011waveform}, a large body of literature has been devoted to the design of communication-centric schemes \cite{9359665,9005192,9109735,yuan2023otfs,FanLiu2024}. Due to the convenience of implementation, early endeavours on this direction performed sensing relying on pilot symbols for channel estimation \cite{80211ad}. To fully unleash the sensing potential of communication signals, recent works further took data payload signals into account \cite{YS2022,FanLiu2024,XiongTIT,giuseppeTIT}. Notably, the authors of \cite{FanLiu2024} have shown that \ac{ofdm} is the optimal communication-centric \ac{isac} waveform for data payload signals, in the sense that it achieves the lowest ranging sidelobe.

Despite their convenience, communication-centric \ac{isac} waveforms in their original form may not exhibit satisfactory long-range sensing capability, which is crucial for applications such as surveillance of low-altitude unmanned aerial vehicles (UAVs). To elaborate, due to the insufficient transmitter-receiver isolation, straightforward utilisation of the communication-oriented signals for sensing would suffer from \acf{si}, thereby exhibiting limited sensing range. In its essence, the \ac{si} issue originates from the fact that the sensing subsystem operates in a \emph{full-duplex} mode, which receives echoes while transmitting communication signals. Such problems have been discussed extensively in the communication literature \cite{fd1,fd2,fd3,fd4} but remain far from being completely resolved.

In the radar literature, a classical solution to \ac{si} is to use pulse radars, which switch off the receiver upon transmission. Such transmission schemes are often referred to as \emph{half-duplex} schemes in the communication literature \cite{hd_isac}. Since echoes returned during transmission cannot be received, the pulses employed in pulse radars are typically narrow to ensure relatively small blind ranges. Consequently, pulse radars designed for ranging typically have small duty cycles ($\leq 10$\%) \cite[Chap.~19]{radar_analysis}. Nevertheless, in the context of \ac{isac}, the requirement of high-throughput communication motivates large duty cycle designs that can carry more data symbols on the pulses. 

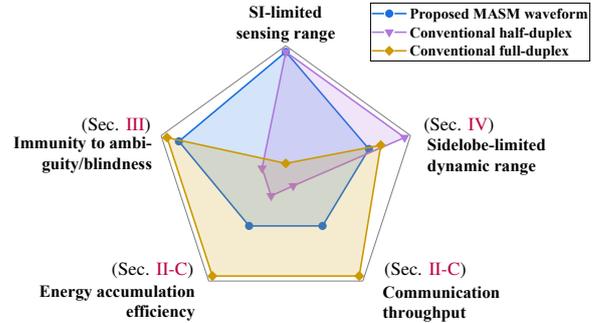
\begin{figure}[t]
    \centering
    \begin{overpic}[width=0.9\linewidth]{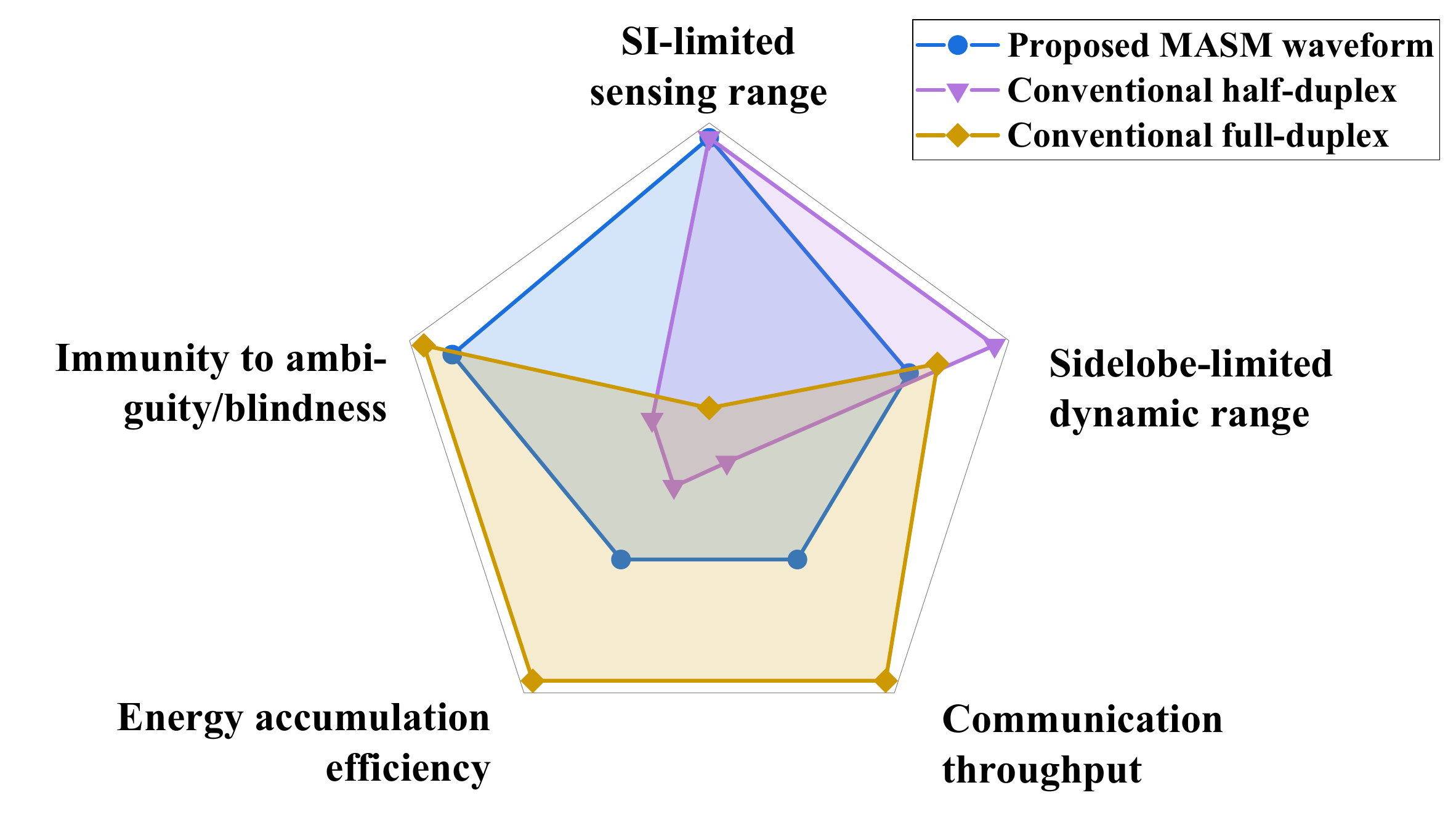}
        \put(20,11){\scriptsize (Sec.~\ref{ssec:metric})}
        \put(65,11){\scriptsize (Sec.~\ref{ssec:metric})}
        \put(15,35){\scriptsize (Sec.~\ref{sec:range_glint})}
        \put(71.5,35){\scriptsize (Sec.~\ref{sec:sidelobe})}
    \end{overpic}
    \caption{Qualitative performance comparison between proposed \ac{masm} waveforms and conventional \ac{isac} waveforms.}
    \label{fig:radarplot_sptm}
    \vspace{-4mm}
\end{figure}

To tackle this sensing-communication tradeoff, a possible approach is to first increase the duty cycle (resulting in higher \ac{prf}), and then mitigate the range ambiguity issue using \ac{prf} staggering \cite{staggerMTI,stagger_new1,stagger_new2,iet_ripple}. The main idea of \ac{prf} staggering methods is to jointly use a set of different \ac{prf}s, favourably being co-prime with one another. Since the length of consecutive \acp{pri} are different, target echoes delayed longer than a single \ac{pri} can be resolved. Besides long-range ranging, \ac{prf} staggering techniques are also applied to mitigate the Doppler ambiguity issue (also known as ``blind speed'') in moving target indicator (MTI) radars \cite{staggerMTI}. The main drawback of \ac{prf} staggering methods is that there is a lack of principled design methods of the stagger ratios, namely the ratio between the \acp{pri}. Furthermore, they typically introduce deleterious fluctuations (``ripples'' \cite{iet_ripple}) to the target response spectrum. In light of this, a systematic waveform design approach striking favourable sensing-communication tradeoffs is highly desirable.

Against the aforementioned background, in this treatise, we propose a half-duplex ISAC waveform design scheme, termed as \acf{masm}. In particular, \ac{masm} designs transmission masks minimizing a range blindness metric, referred to as ``mainlobe fluctuation'', given a communication throughput (or equivalently, duty cycle) constraint. In each \ac{pri}, the system transmits when the mask entry equals to $1$, and receives when it equals to $0$. It is worth highlighting that the proposed design is capable of supporting relatively high-throughput communication ($\sim$50\% duty cycle) under small blind ranges and mild range ambiguity. A qualitative illustration of the performance of \ac{masm} waveforms can be seen in Fig.~\ref{fig:radarplot_sptm}, with conventional half-duplex (i.e., pulse radar) and full-duplex schemes being benchmarks. 

The main contributions of this treatise are:
\begin{itemize}
    \item We characterise the mainlobe fluctuation effect by defining the metrics of \acf{rgi} and \acf{irgi}, and highlight that it is a unique phenomenon concerning half-duplex sensing schemes;
    \item We derive explicit closed-form expressions for the expected \ac{irgi} of conventional pulse radars and \ac{masm}, and formulate the mainlobe fluctuation mitigation problem as an $\ell_4$-norm minimisation problem;
    \item We prove that under certain configurations of mask length $N$ and duty cycle $\rho$, there exist mainlobe-fluctuation-ideal transmission masks yielding zero \ac{irgi} for constant modulus constellations, by \emph{explicit construction};
    \item We analyse the sidelobe performance of \ac{masm}, and prove that there exist transmission masks (including $m$-sequences) that are simultaneously ideal in the sense of mainlobe fluctuation and sidelobe level, by \emph{explicit construction};
    \item We show that \ac{masm} can be applied as frame-level waveforms operating on the slow-time scale, and derive closed-form expressions characterising the relationship between slow-time and fast-time performance metrics.
\end{itemize}

The rest of this treatise is organised as follows. In Sec.~\ref{sec:model}, we present the system model, including the transmit signal model, sensing reception processing and performance metrics. Based on these models, we analyse the mainlobe fluctuation effect and formulate the mainlobe fluctuation mitigation problem in Sec.~\ref{sec:range_glint}. We further analyse the sidelobe performance in Sec.~\ref{sec:sidelobe}, and discuss the application of \ac{masm} as slow-time coding in Sec.~\ref{sec:slow_time}. We verify and demonstrate the analytical results using numerical examples in Sec.~\ref{sec:numerical}, and finally conclude the treatise in Sec.~\ref{sec:conclusion}. 

\subsubsection*{Notations}
The notations $\M{0}_{M\times N}$, $\M{1}_{M\times N}$, $\M{I}_{m}$, $\V{1}_{m}$, and $\V{0}_{m}$ denote the $M\times N$ all-zero matrix, $M\times N$ all-one matrix, identity matrix of order $m$, $m$-dimensional all-one vector and $m$-dimensional all-zero vector, respectively. When they are clear from the context, the subscripts may be omitted. The notations $[\M{A}]_{k,:}$, $[\M{A}]_{:,l}$ and $[\M{A}]_{i,j}$ denote the $k$-th row, the $l$-th column, and the $(i,j)$-th entry of matrix $\M{A}$, respectively. $[\V{x}]_{a:b}$ denotes the vector obtained by extracting the $a$-th to $b$-th entries from $\V{x}$. $[\V{m}]_i$ denotes the $i$-th entry of vector $\V{m}$. $[\V{x};\V{y}]$ and $[\V{x},\V{y}]$ denote vertical and horizontal concatenation of vectors, respectively. The notation $|\V{a}|$ denotes the entrywise magnitude of $\V{a}$, while $|\V{a}|^2$ denotes the entrywise square of $|\V{a}|$. $\M{A}\odot\M{B}$ denotes the Hadamard product. ${\rm diag}(\V{x})$ denotes the diagonal matrix constructed by placing $\V{x}$ on its main diagonal. $\|\cdot\|_p$ denotes the $\ell_p$-norm of its argument, with $p=2$ when the subscript is omitted. $\langle f,g\rangle$ denotes the inner product between $f$ and $g$.

\section{System Model}\label{sec:model}
Let us denote by $T_c$ the symbol interval so that the bandwidth is $\frac{1}{T_c}$, and by $NT_c$ the maximum distance at which we want to undertake sensing. As a consequence, the PRI coincides with $NT_c$. Let us also assume that we process a single \ac{pri} for sensing, so that the number of samples included in the processing interval is $N$. 
\subsection{Transmit Signal Model}
The ISAC transmitter forms a data signal with duty cycle $\rho$, which means that it transmits the signal
\begin{equation}\label{transmit_model}
x(t)=\sum_{i=-\infty}^\infty m_{\rm t}(i)x_i\psi(t-iT_c), 
\end{equation}
where $m_{\rm t}(i)$ is a periodic $(0,1)$-binary sequence with period $N$, referred to as the transmission mask, $\psi(t)$ denotes the pulse shaping filter in communication systems, and $\{x_i\}$ represents a zero-padded data sequence satisfying $x_i=0$ for $m_{\rm t}(i)=0$, while for $m_{\rm t}(i)=1$ it holds that $x_i\in\Set{S}$, with $\Set{S}\subset \mathbb{C}$ being the coding alphabet (or constellation set). We assume that the communication symbols $x_i\neq 0$ are drawn from $\Set{S}$ in an \ac{iid} manner, satisfying the following assumptions.
\begin{assumption}[Unit Power]\label{asu:unit_power}
    We consider constellations having unit power, namely, 
    \begin{equation}\label{unit_power}
        \mathbb{E}(|x_i|^2) = 1,\quad \forall i.
    \end{equation}
\end{assumption}
\noindent Assumption \ref{asu:unit_power} normalises the power of the constellations, ensuring that they can be fairly compared against each other in terms of their sensing and communication performance.
\begin{assumption}[Rotational Symmetry]\label{asu:rotsym}
    We consider constellation having zero expectation and zero pseudo-variance, namely,
    \begin{equation}
        \mathbb{E}(x_i) = 0,\quad\mathbb{E}(x_i^2) = 0, \quad \forall i.
    \end{equation}
\end{assumption}
\subsection{Sensing Reception Processing Model}
\begin{figure}[t]
    \centering
    \includegraphics[width=0.78\linewidth]{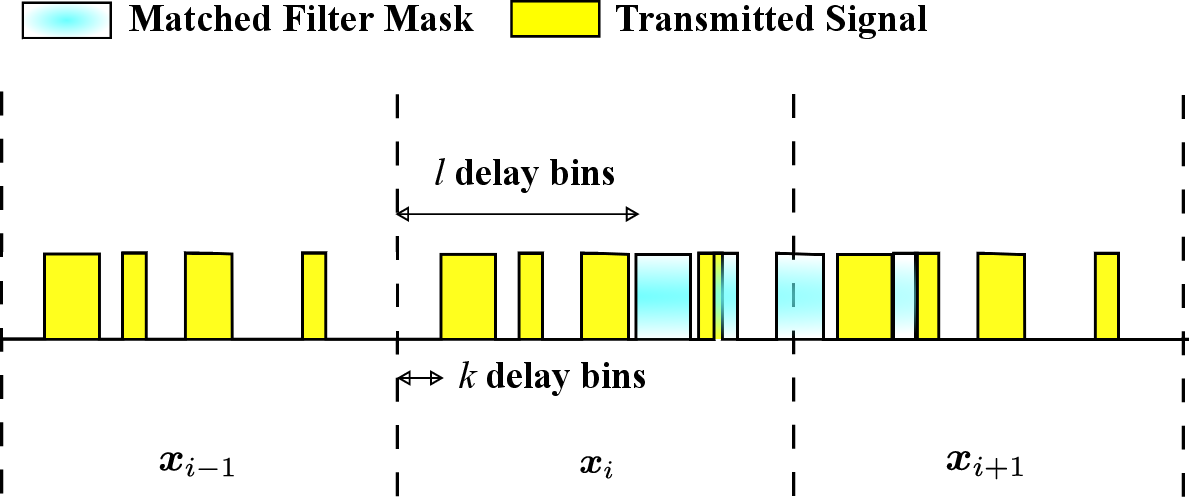}
    \caption{Illustration of the reception processing in \eqref{range_profile}. We focus on the $0$-th \ac{pri} $\V{x}_0$ throughout the treatise, although all results also apply to other \ac{pri}s.}
    \label{fig:three_intervals}
    \vspace{-4mm}
\end{figure}

If a single target is located at distance $R$ from the transmitter, the transmitted signal would be reflected back to the receiver in a delayed manner, with the delay being $\tau_0=\frac{2R}{c}$, whereby the noiseless signal impinging on the receive antenna would be
\[
z_0(t)=x(t-\tau_0)=\sum_{i=-\infty}^\infty m_{\rm t}(i)x_i\psi(t-\tau_0-iT_c),
\]
in keeping with \eqref{transmit_model}. We do not consider the additive noise since it does not alter the reception process. The receiver may operate only during certain symbol intervals. To this end, we define the reception mask $m_{\rm r}(i)$, which is also a periodic $(0,1)$-binary sequence with period $N$. The receiver undertakes the projections onto the basis 
$\{m_{\rm r}(n)\psi(t-nT_c)\}$, thus yielding the observations
\begin{equation}\label{mf_conventional}
z_0(n)=\langle z(t), m_{\rm r}(n)\psi(t-nT_c)\rangle.  
\end{equation}
For full-duplex systems, \eqref{mf_conventional} is equivalent to the conventional matched filtering. For half-duplex systems, we have $m_{\rm r}(i)=1-m_{\rm t}(i)$, thereby the projection yields
$$
z_0(n)=m_{\rm r}(n)\sum_{i=-\infty}^\infty x_im_{\rm t}(i)\langle\psi(t-\tau_0-iT_c),\psi(t-nT_c)\rangle.
$$
If $\psi (\cdot )$ satisfies the Nyquist criterion with respect to $T_c$, and assuming $\tau_0=kT_c$, then we obtain the sampled signal
\[
z_k(n)=m_{\rm r}(n)m_{\rm t}(n-k)x_{n-k} \; , n \in \mathbb{Z}.
\]
Since the processing interval is $N$, we actually have to process the finite-length sequence
\[
\overline{z}_k(n)=z_k(n)R_{N}(n) \; ,~ R_{Q}(n)= \left\{
\begin{array}{ll}
1 & 0 \leq n \leq Q-1\\
0 & \mbox{otherwise}
\end{array} \right.
\]
The transmitted sequence, on the other hand, reads
\[
\overline{z}(n)=m_{\rm t}(n)x_{n}
\]
whereby the output of the correlator reads
\begin{align}
r_{k,l}&=\overline{z}_k(n)\circledast\overline{z}^*(l-n) \nonumber\\
&=\sum_{n=-\infty}^\infty m_{\rm r}(n)m_{\rm t}(n-k)
m_{\rm t}(n-l)x_{n-k}x^*_{n-l}R_{N}(n),\nonumber
\end{align}
namely
\begin{equation}\label{mf_output}
r_{k,l}=\sum_{n=0}^{N-1} m_{\rm r}(n)m_{\rm t}(n-k)
m_{\rm t}(n-l)x_{n-k}x^*_{n-l}.
\end{equation}
Whenever it is more convenient, we may use the following vectorized notations:
\begin{subequations}
\begin{align}
\V{m}_{\rm t}&=[m_{\rm t}(0),\dotsc,m_{\rm t}(N-1)]^{\rm T},\\
\V{m}_{\rm r}&=[m_{\rm r}(0),\dotsc,m_{\rm r}(N-1)]^{\rm T}, \\
\V{x}_i &= [x_{iN},\dotsc,x_{(i+1)N-1}]^{\rm T},
\end{align}
\end{subequations}
which amount to
\begin{align}\label{range_profile}
r_{k,l} &= [\V{0}_N;\V{x}_0;\V{0}_N]^{\rm H}(\M{I}_3\otimes \M{M}_{\rm t})\widetilde{\M{J}}_l^{\rm H} (\M{I}_3\otimes \M{M}_{\rm r})\widetilde{\M{J}}_k\nonumber\\
&\hspace{5mm}\cdot(\M{I}_3\otimes\M{M}_{\rm t}) [\V{x}_{-1};\V{x}_0;\V{x}_1],
\end{align}
as portrayed in Fig.~\ref{fig:three_intervals}, where $\M{M}_{\rm t}={\rm diag}(\V{m}_{\rm t})$, $\M{M}_{\rm r}={\rm diag}(\V{m}_{\rm r})$, and $\widetilde{\M{J}}_k$ denotes the $n$-th periodic shift matrix taking the following form
\begin{equation}
    \widetilde{\M{J}}_k = \left[ {\begin{array}{*{20}{c}}
  {\M{0}}& {\M{I}_{k\times k}}\\ 
  {{\M{I}_{(3N-k)\times (3N-k)}}}&{\M{0}} 
\end{array}} \right].
\end{equation}
We shall denote the $i$-th entry in $\V{m}_{\rm t}$ and $\V{m}_{\rm r}$ by $m_{{\rm t},i}$ and $m_{{\rm r},i}$, respectively. The duty cycle can then be expressed as 
\begin{equation}\label{dc}
\rho = \frac{1}{N}\V{1}^{\rm T} \V{m}_{\rm t}.
\end{equation}
\subsection{Performance Metrics}\label{ssec:metric}
The sensing and communication performance may be evaluated using metrics derived from the transmission mask and the corresponding range response. Next, we discuss several important performance metrics.

\subsubsection{Mainlobe fluctuation}
The mainlobe level is often of interest, given by
\begin{equation}\label{mainlobe}
r_{k,k}=\sum_{n=0}^{N-1} m_{\rm r}(n)m_{\rm t}(n-k)|x_{n-k}|^2,
\end{equation}
which depicts the total received energy of an echo within the $k$-th delay bin. In the full-duplex case, the mainlobe level can simply be obtained as
$$
r_{k,k}= \V{x}_0^{\rm H} \widetilde{\M{J}}_k^{\rm H}\widetilde{\M{J}}_k\V{x}_0 = \|\V{x}_0\|_2^2,
$$
which is constant with respect to $k$. Unfortunately, this is no longer the case for half-duplex systems. Such fluctuations of mainlobe levels across different delay bins would result in unexpected misdetections or false alarms, especially in multi-target sensing scenarios. In the rest of this treatise, we will refer to this phenomenon as the ``mainlobe fluctuation'' effect of half-duplex systems. 

Formally, we may define the \acf{rgi} of the $k$-th range bin as
\begin{equation}\label{rgi_def}
g_k(\V{r}_{\rm m}) = \left|r_{k,k} - \frac{1}{N-1} \sum_{l=1}^{N-1} r_{l,l}\right|^2,
\end{equation}
for $k=1,\dotsc,N-1$, where $\V{r}_{\rm m}=[r_{0,0},\dotsc,r_{N-1,N-1}]^{\rm T}$. Note that we do not consider the \ac{rgi} at $k=0$, since $r_{0,0}=0$ holds for all half-duplex sensing systems, and hence any target within this range bin is undetectable (known as the blind range of half-duplex sensing systems). When the global mainlobe fluctuation performance is considered, we may also define the \acf{irgi} as follows
\begin{equation}\label{argi_definition}
\overline{g}(\V{r}_{\rm m}) = \frac{1}{N-1}\sum_{k=1}^{N-1} |r_{k,k}|^2 - \left|\frac{1}{N-1} \sum_{l=1}^{N-1} r_{l,l}\right|^2.
\end{equation}
A sensing signal with a low \ac{irgi} is thus desired for half-duplex sensing systems. For \ac{isac} systems, since the communication symbols $\V{x}_i$ are random, one may wish to use the expected \ac{irgi}
$$
{\rm EARGI} = \mathbb{E}_{\V{x}_0}\{\overline{g}(\V{r}_{\rm m})\}
$$
to evaluate the performance of the mainlobe of the range response. We observe that the expected \ac{irgi} is in fact independent of the communication symbols, and hence applies to all \ac{pri}s (not limited to $\V{x}_0$).
\begin{remark}[ARGI as Conditional Variance]
At this stage, let us elaborate on the physical implications of the aforementioned mainlobe fluctuation metrics. In particular, the \ac{irgi} may be interpreted as the variance of mainlobe level across delay bins $k>1$, given by
$$
\overline{g}(\V{r}_{\rm m}) = \frac{1}{N-1}\sum_{k=1}^{N-1} \left|r_{k,k} - \frac{1}{N-1}\sum_{l=1}^{N-1} r_{l,l}\right|^2,
$$
conditioned on a fixed realization of the communication symbols $\V{x}_0$. The \ac{rgi} is thus a metric depicting the deviation of $r_{k,k}$ to the mean mainlobe level across all delay bins.
\end{remark}

Note that \ac{irgi} (and hence expected \ac{irgi}) has the following properties:
\begin{proposition}\label{prop:irgi_property}
It follows that
    \begin{enumerate}
    \item $\overline{g}(\V{r})=\overline{g}(-\V{r})$;
    \item $\overline{g}(\V{r}+c\V{1})=\overline{g}(\V{r}),~\forall c\in\mathbb{R}$.
\end{enumerate}
\end{proposition}

\subsubsection{Sidelobe level metrics}
The (squared) sidelobe levels $|r_{k,l}|^2,~k\neq l$, are also useful. In particular, a widely applied performance criterion, the \ac{isl}, is defined based on the sidelobe levels as follows
\begin{equation}\label{isl}
{\rm ISL}_k = \sum_{l\neq k}|r_{k,l}|^2,
\end{equation}
where is independent of $k$ for full-duplex systems. For conventional pulse radars, the \ac{isl} is also independent of $k$ outside their blind ranges. For \ac{masm} systems, \ac{isl} is no longer the most suitable sidelobe performance metric, since it not only relies on $k$ but also depends on the random data payload. To this end, in this treatise, we consider the \ac{aesl}, which is averaged across all range bins, defined as
\begin{equation}
{\rm AESL} = \frac{1}{(N-1)(N-2)}\sum_{k=1}^{N-1} \sum_{l>0,k\neq k} \mathbb{E}\{|r_{k,l}|^2\}.
\end{equation}
In some scenarios, the peak sidelobe level is also useful. Thus we define the \ac{pesl} as follows
\begin{equation}
{\rm PESL} = \max_{k>0,l>0,l\neq k} \mathbb{E}\{|r_{k,l}|^2\}.
\end{equation}

\subsubsection{Energy accumulation efficiency}
Since full-duplex systems transmit continuous waveforms, they typically accumulate echo energy more efficiently than half-duplex systems. Following this intuition, we define the energy accumulation efficiency as the ratio between the actual received echo energy and the received echo energy of a full-duplex system having identical configurations (including constellations, transmit power, frequency band, etc.) except for the transmission and reception masks, averaged across all delay bins, given by
\begin{equation}
\eta_{\rm S} = \frac{\sum_{k=0}^{N-1}\mathbb{E}\{r_{k,k}\}}{N \mathbb{E}\{\|\V{x}_i\|_2^2\}}=\frac{1}{N^2}\sum_{k=0}^{N-1}\mathbb{E}\{r_{k,k}\}.
\end{equation}
Naturally, the energy accumulation efficiency of full-duplex systems is $1$. For half-duplex systems, using the aforementioned assumptions \eqref{dc} and \eqref{unit_power}, we have
\begin{align}\label{expected_mainlobe}
\mathbb{E}\{r_{k,k}\}&=\mathbb{E}\left\{\left\|(\M{I}-\M{M}_{\rm t})\widetilde{\M{J}}_k\M{M}_{\rm t}\V{x}_i\right\|_2^2\right\} \nonumber\\
&= (\V{1}-\V{m}_{\rm t})^{\rm T}\left(\widetilde{\M{J}}_k(\V{m}_{\rm t}\odot \mathbb{E}\left\{|\V{x}_i|^2\right\})\right)\nonumber\\
&=\rho N - \V{m}_{\rm t}^{\rm T}\widetilde{\M{J}}_k\V{m}_{\rm t}.
\end{align}
By rewriting the correlation $\{\V{m}_{\rm t}^{\rm T}\widetilde{\M{J}}_k\V{m}_{\rm t}\}_{k=0}^{N-1}$ in the frequency domain, we further obtain
\begin{align}
\mathbb{E}\{\V{r}_{\rm m}\} &= \rho N\V{1} - \sqrt{N}\M{F}^{\rm H}|\M{F}\V{m}_{\rm t}|^2,
\end{align}
where
$$
\M{F}=\frac{1}{\sqrt{N}}\left[ {\begin{array}{*{20}{c}}
\omega_N^{0\cdot 0} & \omega_N^{0\cdot 1} & \dotsc & \omega_N^{0\cdot (N-1)}\\
\omega_N^{1\cdot 0} & \omega_N^{1\cdot 1} & \dotsc & \omega_N^{1\cdot (N-1)}\\
\vdots & \vdots & \ddots & \vdots \\
\omega_N^{(N-1)\cdot 0} & \omega_N^{(N-1)\cdot 1} & \dotsc & \omega_N^{(N-1)\cdot (N-1)}
\end{array}}\right]
$$
denotes the unitary matrix representing the discrete Fourier transform, with $\omega_N=\exp(-2\pi i/N)$. Therefore
$$
\begin{aligned}
\V{1}^{\rm T}\mathbb{E}\{\V{r}_{\rm m}\}&=\rho N^2 - \sqrt{N}\V{1}^{\rm T}\M{F}^{\rm H}|\M{F}\V{m}_{\rm t}|^2\\
&=\rho N^2 - N [1; \V{0}_{N-1}]^{\rm T}|\M{F}\V{m}_{\rm t}|^2\\
&= \rho N^2 - N |\V{f}_1^{\rm T}\V{m}_{\rm t}|^2\\
&= \rho N^2 - \rho^2 N^2.
\end{aligned}
$$
where $\V{f}_1^{\rm T}=\frac{1}{\sqrt{N}}\V{1}^{\rm T}$ is the first row of $\M{F}$. We may then conclude that
$$
\eta_{\rm S} = \frac{\rho N^2 - \rho^2 N^2}{N^2} = \rho(1-\rho).
$$
It is clear that when $\rho=0.5$, the energy accumulation efficiency achieves its maximum value $0.25$. Systems with higher energy accumulation efficiency can achieve a larger \ac{snr} under the same peak power constraint.

\subsubsection{Communication throughput}
From the perspective of the communication system, the \ac{isac} signal should convey as much information as possible. The communication throughput of \ac{masm} systems can be computed as
\begin{equation}
T = \frac{1}{N} \V{1}^{\rm T} \V{m}_{\rm t} \log_2(|\Set{S}|) = \rho \log_2(|\Set{S}|)
\end{equation}
for each channel use \cite{throughput}, where $\rho$ is the duty cycle and $|\Set{S}|$ represents the order of the constellation. In light of this, given a fixed constellation, the communication throughput constraint can be simplified as a duty cycle constraint.

\section{The Range Glint Effect: Analysis and Mitigation}\label{sec:range_glint}
\begin{figure}[t]
    \centering
    \includegraphics[width=0.85\linewidth]{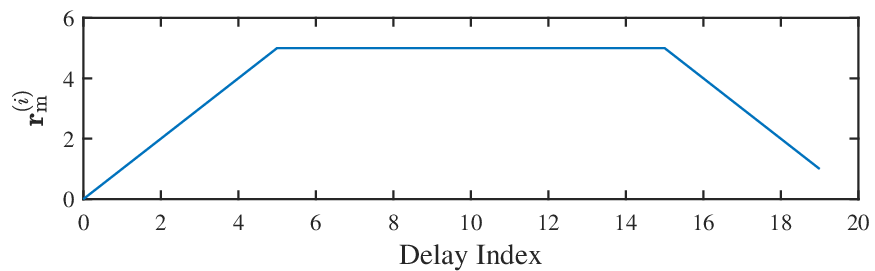}\\
    \includegraphics[width=0.85\linewidth]{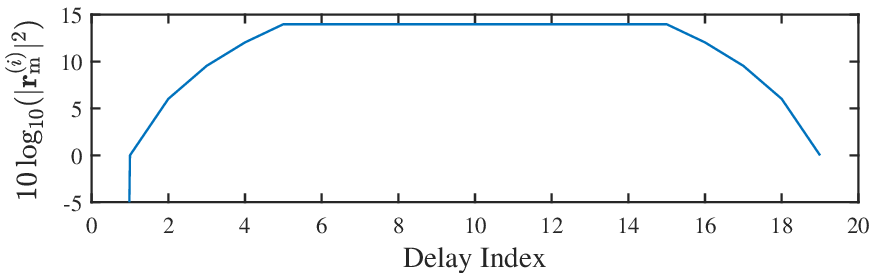}
    \caption{Illustration of the mainlobe fluctuation effect of conventional pulse radars. Trapezoidal $\V{r}_{\rm m}$'s suffer from severe mainlobe fluctuation across many range bins.}
    \label{fig:trapezoid}
    \vspace{-4mm}
\end{figure}
\subsection{Range Glint of Conventional Pulse Radars}\label{ssec:conventional}
Let us commence with a mainlobe fluctuation analysis of conventional pulse radar waveforms, which will in turn motivate the design of \ac{masm}. A typical pulse radar has a continuous transmission interval in each \ac{pri}, namely
\begin{equation}\label{continuous_mask}
\V{m}_{\rm t} = [\V{1}_{\rho N};~\V{0}_{(1-\rho)N}].
\end{equation}
The waveforms of pulse radars are typically constant modulus, deterministic signals, with linear frequency modulated signals being a representative example. For these waveforms, according to \eqref{expected_mainlobe}, the mainlobe of the range response can be written as
\begin{align}
r_{k,k} = \mathbb{E}\{r_{k,k}\} = (\V{1}-\V{m}_{\rm t})^{\rm T}\widetilde{\M{J}}_k \V{m}_{\rm t}.
\end{align}
It then follows that
\begin{equation}
\V{r}_{\rm m} =\rho N\V{1} - \sqrt{N}\M{F}^{\rm H}|\M{F}\V{m}_{\rm t}|^2.
\end{equation}
Note that $\sqrt{N}\M{F}^{\rm H}|\M{F}\V{m}_{\rm t}|^2$ is in fact the periodical \ac{acf} of $\V{m}_{\rm t}$. This suggests that the significance of the mainlobe fluctuation effect may be intuitively depicted by the flatness of the \ac{acf}. For continuous transmission masks taking the form of \eqref{continuous_mask}, the periodical \ac{acf} is the convolution of a boxcar function with itself, and hence has a triangular shape. Consequently, as a function of $m$, $\V{r}_{\rm m}$ has a trapezoidal shape, as illustrated in Fig.~\ref{fig:trapezoid}. The regions (especially the one containing the zeroth delay bin) in which $\V{r}_{\rm m}$ does not take its maximum value $\rho N$ are typically referred to as the \emph{blind range} in the radar literature \cite[Sec.~1.5]{radar_analysis}. In the context of the mainlobe fluctuation effect, the blind range may be defined as a set $\Set{B}$ of delay bins in which the system suffers from severe mainlobe fluctuation, in the sense that
$$
|r_{k,k}|^2\leq \frac{c_{\rm th}}{N}\sum_{l=1}^N |r_{l,l}|^2,~\forall k\in\Set{B},
$$
where $c_{\rm th}<1$ is a threshold depicting the system's tolerance of mainlobe fluctuation.  Apparently, for conventional pulse radars, the blind range expands as the duty cycle $\rho$ increases, thereby raising a tradeoff between the target sensing capability and communication throughput.

\subsection{Range Glint Mitigation Problem}
Naturally, we would expect that a more favourable sensing-communication tradeoff can be achievable by using transmission masks suffering from milder mainlobe fluctuation, or quantitatively, having a smaller \ac{irgi}. This requirement may be formulated as the following optimisation problem
\begin{subequations}\label{opt_cm}
\begin{align}
    \min_{\V{m}_{\rm t}\in\{0,1\}^N}&~~\overline{g}(\V{r}_{\rm m}),\\
    {\rm s.t.}&~~\V{1}^{\rm T}\V{m}_{\rm t}=\rho N.
\end{align}
\end{subequations}
\subsubsection{Deterministic Constant Modulus Waveforms}
For pulse radars employing deterministic, constant modulus signals, we have the following result.
\begin{proposition}\label{prop:det_irgi}
The \ac{irgi} of determinisitic constant modulus signals can be computed as 
\begin{equation}
\overline{g}(\V{r}_{\rm m}) =\frac{N}{N-1}\|\M{F}\V{m}_{\rm t}\|_4^4-\frac{\rho^2 N^3\left(\rho^2N-2\rho+1\right)}{(N-1)^2}.
\end{equation}
\begin{IEEEproof}
Please refer to Appendix \ref{sec:proof_det_irgi}.
\end{IEEEproof}
\end{proposition}

Proposition \ref{prop:det_irgi} implies that the objective function in the optimisation problem \eqref{opt_cm} can be replaced with the simpler term $\|\M{F}\V{m}_{\rm t}\|_4^4$. Note that $\|\M{F}\V{m}_{\rm t}\|_4^4 = \left\||\M{F}\V{m}_{\rm t}|^2\right\|_2^2$, where $|\M{F}\V{m}_{\rm t}|^2$ is the power spectrum of $\V{m}_{\rm t}$. A closely related quantity is the power spectral variance of $\V{m}_{\rm t}$ given by
\begin{equation}\label{psv}
{\rm PSV}(\V{m}_{\rm t}) := \frac{1}{N}\V{1}^{\rm T} \left(|\M{F}\V{m}_{\rm t}|^2-\frac{1}{N}\V{1}^{\rm T}|\M{F}\V{m}_{\rm t}|^2\right)^2,
\end{equation}
which can be rewritten as
$$
{\rm PSV}(\V{m}_{\rm t}) = \frac{1}{N}\|\M{F}\V{m}_{\rm t}\|_4^4-\rho^2.
$$
Thus we may now conclude that designing constant modulus low-\ac{irgi} transmission masks is equivalent to minimizing the power spectral variance of the mask. In other words, transmission masks having flat power spectra will suffer less from the mainlobe fluctuation effect.

\subsubsection{Data Payload Signals}
\begin{table*}[t]
\caption{The scaling of the ratio between the expected \ac{irgi} and the average expected mainlobe level (averaged over delay bins), with respect to $N$, for different data payload signals.}
    \label{tab:eirgi_scaling}
    \centering
    \begin{tabular}{|c|c|c|} \hline
         & Constant modulus constellations & Generic constellations \\ \hline
        Conventional pulse & $O(1)$ & $O(1)$ \\ \hline
        Random transmission masks & $O(1/N)$ & $O(1/N)$ \\ \hline
        Mainlobe-fluctuation-ideal transmission masks & 0 & $O(1/N)$ \\ \hline
    \end{tabular}
    \vspace{-4mm}
\end{table*}
In general, for \ac{isac} systems, we have to account for non-constant modulus transmit signals, modulated by random data payload via the orthonormal transformation $\M{U}$. Consequently, the expected \ac{irgi} would depend on the constellations of the data payload signal, detailed as follows.

\begin{proposition}[Expected \ac{irgi} of data payload signals]\label{prop:eirgi}
The expected \ac{irgi} of data payload signals can be expressed as
\begin{align}\label{eirgi_data}
\mathbb{E}_{\V{x}_i}\{\overline{g}(\V{r}_{\rm m})\}
&=\frac{N}{N-1}\|\M{F}\V{m}_{\rm t}\|_4^4-\frac{\rho^2 N^3}{(N-1)^2}(\rho^2 N -2\rho +1) \nonumber \\
&\hspace{3mm}+(\mu_4-1)\cdot \frac{\rho(1-\rho) N^2}{(N-1)^2}(\rho N-1).
\end{align}
\begin{IEEEproof}
Please refer to Appendix \ref{sec:proof_eirgi}.
\end{IEEEproof}
\end{proposition}

We may now conclude that for data payload signals with generic constellations, the objective function in \eqref{opt_cm} can be replaced with the $\ell_4$ norm $\|\M{F}\V{m}_{\rm t}\|_4$, since the constellation-dependent term is independent of $\V{m}_{\rm t}$.
\begin{remark}
Since $1-\rho \geq 0$ and $\rho N-1 \geq 0$ hold for all transmission masks, we see that the expected \ac{irgi} increases with $\mu_4$, with constant modulus constellations being the optimum. This suggests a \ac{drt}: sensing tasks favor constant modulus constellations, whereas the communication task would favor constellations having larger $\mu_4$ for more efficient information transmission.
\end{remark}
\subsection{Solutions to the mainlobe fluctuation mitigation problem}\label{ssec:solving_rgm}
The mainlobe fluctuation mitigation problem can now be written as
\begin{subequations}\label{opt_generic}
\begin{align}
    \min_{\V{m}_{\rm t}\in\{0,1\}^N}&~~\|\M{F}\V{m}_{\rm t}\|_4^4,\\
    {\rm s.t.}&~~\V{1}^{\rm T}\V{m}_{\rm t}=\rho N.
\end{align}
\end{subequations}
This is a nonlinear integer programming problem that typically does not admit efficient algorithms yielding exact solutions. In the range of $N<30$, it is possible to obtain the optimal solution by means of an exhaustive search. For large-$N$ scenarios, in general, one may obtain suboptimal solutions by using the branch and bound approach \cite{bnb}, which can be relatively efficient in this context since the objective function $\|\M{F}\V{m}_{\rm t}\|_4^4$ is convex in $\V{m}_{\rm t}$. 

Fortunately, we may obtain optimal solutions under certain configurations of $\rho$ and $N$. In particular, these solutions are not only optimal, but also \emph{ideal} in the sense that their expected \ac{irgi} can be exactly zero for constant modulus constellations, and hence can be computed as follows:
\begin{align}\label{eirgi_cds}
\mathbb{E}_{\V{x}_0}\{\overline{g}(\V{r}_{\rm m})\} &=(\mu_4-1)\cdot \frac{\rho(1-\rho) N^2}{(N-1)^2} (\rho N-1).
\end{align}
In what follows, we shall refer to these $\V{m}_{\rm t}$'s as \emph{mainlobe-fluctuation-ideal transmission masks}, and refer to the corresponding data payload signals as \emph{mainlobe-fluctuation-ideal data payload signals}. Let us denote
$$
a_k := \mathbb{E}\{r_{k,k}\}=(\V{1}-\V{m}_{\rm t})^{\rm T}\widetilde{\M{J}}_k\V{m}_{\rm t},
$$
and $\V{a}=[a_0,\dotsc,a_{N-1}]^{\rm T}$, we have the following result.

\begin{lemma}\label{lem:eargi_alternative}
It holds that
\begin{align}\label{time_domain_eirgi}
&\frac{N}{N-1}\|\M{F}\V{m}_{\rm t}\|_4^4-\frac{\rho^2 N^3}{(N-1)^2}(\rho^2 N -2\rho +1)\nonumber \\
& \hspace{3mm}= \frac{1}{N-1}\sum_{k=1}^{N-1} \left(a_k-\frac{1}{N-1}\sum_{l=1}^{N-1}a_l\right)^2.
\end{align}
\begin{IEEEproof}
According to Proposition \ref{prop:det_irgi}, we have
\begin{equation}\label{alternative1}
\overline{g}(\V{a}) = \frac{N}{N-1}\|\M{F}\V{m}_{\rm t}\|_4^4-\frac{\rho^2 N^3}{(N-1)^2}(\rho^2 N -2\rho +1).
\end{equation}
But from \eqref{argi_definition}, we obtain
\begin{align}\label{alternative2}
\overline{g}(\V{a}) &=\frac{1}{N-1}\sum_{k=1}^{N-1} a_k^2 - \left(\frac{1}{N-1} \sum_{l=1}^{N-1} a_l\right)^2 \nonumber\\
&=\frac{1}{N-1}\sum_{k=1}^{N-1} \left(a_k-\frac{1}{N-1}\sum_{l=1}^{N-1}a_l\right)^2.
\end{align}
Combining \eqref{alternative1} with \eqref{alternative2} we obtain \eqref{time_domain_eirgi}.
\end{IEEEproof}
\end{lemma}

For conventional continuous pulses, the right-hand side of \eqref{time_domain_eirgi} is on the order of $O(N^2)$. For random transmission masks drawn from $\{0,1\}^N$, we have the following result.

\begin{proposition}[Expected \ac{irgi} of random masks]\label{prop:random_masks}
For random transmission masks whose entries are sampled from \ac{iid} Bernoulli distributions ${\rm Bern}(\rho)$, the expected \ac{irgi} (over both transmission masks and communication symbols) is given by
\begin{align}\label{eirgi_data2}
\mathbb{E}\{\overline{g}(\V{r}_{\rm m})\}
&=\frac{\rho^2(1-\rho)^2 N(N-2)}{N-1} \nonumber \\
&\hspace{3mm}+(\mu_4-1)\cdot \frac{\rho(1-\rho) N^2}{(N-1)^2}(\rho N-1).
\end{align}
\begin{IEEEproof}
Please refer to Appendix \ref{sec:proof_random_masks}.
\end{IEEEproof}
\end{proposition}

Now note that the average expected mainlobe level is in the order of $O(N^2)$. This suggests that as $N$ increases, data payload signals modulated on conventional pulses would have a roughly constant \ac{irgi}-to-mainlobe-level ratio, but for random transmission masks and mainlobe-fluctuation-ideal data payload signals this ratio will be vanishing. Our observations are summarised in Table~\ref{tab:eirgi_scaling}.

To find mainlobe-fluctuation-ideal solutions, let us first note a simple relationship between the periodic \ac{acf} of $(0,1)$-binary sequences and $(+1,-1)$-binary sequences:
\begin{lemma}\label{lem:zo2pm}
Denote the periodic \ac{acf} of a $(0,1)$-binary sequence $\V{x}$ as $\V{r}_{\V{x}}$. Using the transformation $\V{y} = \V{1}-2\V{x}$ one may obtain a $(+1,-1)$-binary sequence. It turns out that the periodic \ac{acf} $\V{r}_{\V{y}}$ of $\V{y}$ satisfies
\begin{equation}\label{zo2pm}
\V{r}_{\V{y}} = N(1-4\rho) + 4\V{r}_{\V{x}}.
\end{equation}
\begin{IEEEproof}
The entries of $\V{r}_{\V{y}}$ can be computed as
$$
\begin{aligned}
\relax[\V{r}_{\V{y}}]_k &=(\V{1}-2\V{x})^{\rm T}\widetilde{\M{J}}_{k-1}(\V{1}-2\V{x})\\
&=N + 4\V{x}^{\rm T}\widetilde{\M{J}}_{k-1} \V{x} - 4 \V{x}^{\rm T}\widetilde{\M{J}}_{k-1}\V{1}\\
&= N - 4 \rho N + 4[\V{r}_{\V{x}}]_k,
\end{aligned}
$$
where $[\V{r}_{\V{x}}]_k$ denotes the $k$-th entry of $\V{r}_{\V{x}}$, thus yielding \eqref{zo2pm}.
\end{IEEEproof}
\end{lemma}

Lemma \ref{lem:zo2pm} implies that the \ac{irgi} of a $(0,1)$-binary sequence (e.g., $\V{m}_{\rm t}$) equals to that of the corresponding $(+1,-1)$-binary sequence. In light of this, we may find mainlobe-fluctuation-ideal transmission masks by resorting to the classical theory of $(+1,-1)$-binary sequence design aiming for finding sequences associated with so-called ``two-level \ac{acf}s'', whose \ac{acf} sidelobes are identical \cite{two_level}.\footnote{Please be aware that the ``sidelobe'' in this context is the sidelobe of the \ac{acf}, rather than the sidelobe levels $|r_{k,l}|^2,~k\neq l$ of the range response.} From the definition of \ac{rgi} \eqref{rgi_def}, it is clear that transmission masks $\V{m}_{\rm t}$ having two-level \ac{acf}s would have zero \ac{rgi} for all $k>0$. A well-known class of sequences belonging to this kind is the $m$-sequences, whose \ac{acf} takes the value of $N$ at the mainlobe, with all sidelobes being equal to $-1$. The Barker codes widely used in the radar literature also satisfy the ``two-level \ac{acf}'' condition. However, all known Barker codes are rather short ($N\leq 13$), making them less suitable for our purpose.

In general, binary sequences having two-level \ac{acf}s are known to be equivalent to the \emph{\acp{cds}} over finite abelian groups \cite{cds}. Formally, a $(\nu,k,\lambda)$-\ac{cds} is a subset $\Set{C}$ of $\mathbb{Z}_\nu$,  satisfying $|\Set{C}|=k$ and
$$
|(w\oplus\Set{C})\cap \Set{C}|=\lambda,~\forall w\in\mathbb{Z}_\nu,
$$
where $w\oplus\Set{C}:=\{(w+n)~{\rm mod}~\nu|n\in\Set{C}\}$. In the context of transmission mask design, we have $\nu = N$, $k=\rho N$, and $\lambda = \rho N - r_{l,l},~\forall l$. A mainlobe-fluctuation-ideal transmission mask $\V{m}_{\rm t}\in\{0,1\}^N$ can be characterised using its associated \ac{cds} $\Set{C}$ in the following manner
\begin{equation}
m_{{\rm t},i} = \left\{
  \begin{array}{ll}
    1, & \hbox{$i-1\in\Set{C}$;} \\
    0, & \hbox{$i-1\notin \Set{C}$.}
  \end{array}
\right.
\end{equation}
A large number of \acp{cds} has been found, as summarised in \cite{lajolla}. Some notable subclasses include GMW \acp{cds} \cite{gmw_sequence}, Paley \acp{cds} \cite{paley_pds}, and Singer \acp{cds} \cite{singer1938theorem}. In particular, Singer \acp{cds} is especially suitable for our purpose, as detailed later in Sec.~\ref{sec:sidelobe}.

\section{Sidelobe Analysis}\label{sec:sidelobe}
In this section, we consider the sidelobe of the range response. 

\subsection{Average Expected Sidelobe}
For unmodulated masks (i.e., satisfying $\V{x}_i \propto \V{1}$), we may construct a matrix $\M{R}$ given by
\begin{equation}
\M{R} \!=\! N\M{F}^{\rm H}{\rm diag}(\M{F}^*\V{m}_{\rm t}^*)\M{F}(\M{I}\!-\!\M{M}_{\rm t})\M{F}^{\rm H}{\rm diag}(\M{F}\V{m}_{\rm t})\M{F},
\end{equation}
which satisfies $[\M{R}]_{k,l} = r_{k,l}$, and thus the off-diagonal entries of $\M{R}$ correspond to the sidelobes, while diagonal entries correspond to the mainlobes at different delay bins. For data payload signals, we have the following result.
\begin{proposition}\label{prop:esl}
The expected sidelobe levels of data payload signals satisfy
\begin{equation}\label{esl}
\mathbb{E}\{|r_{k,l}|^2\} =[\M{R}]_{k,l},~\forall k\neq l.
\end{equation}
\begin{IEEEproof}
Please refer to Appendix \ref{sec:proof_esl}.
\end{IEEEproof}
\end{proposition}

Proposition \ref{prop:esl} suggests a quadratic sidelobe level reduction of random data payload signals over pure masks. Next, we present a simple but important characteristic of the \ac{aesl}.
\begin{proposition}[Constant \ac{aesl}]\label{prop:eavsl}
Given the parameters $N$ and $\rho$, the \ac{aesl} of data payload signals is \emph{independent} of the transmission mask $\V{m}_{\rm t}$. In particular, we have
\begin{equation}\label{eavsl}
\frac{\sum_{k=1}^{N-1}\sum_{l>0,l\neq k} \mathbb{E}\{|r_{k,l}|^2\}}{(N-1)(N-2)}  = \frac{\rho(1-\rho)N^2(\rho N-1)}{(N-1)(N-2)},
\end{equation}
where the average is taken over all sidelobes besides those in the blind range (i.e., $k=0$ or $l=0$).
\begin{IEEEproof}
Please refer to Appendix \ref{sec:proof_eavsl}.
\end{IEEEproof}
\end{proposition}

\subsection{Peak Expected Sidelobe}
\begin{table*}[t]
\caption{Examples of ideal transmission masks associated with Singer \acp{cds}.}
    \label{tab:singer_examples}
    \centering
    \begin{tabular}{|c|c|c|c|c|} \hline
        Projective Space & $N$ & $\rho$ & \ac{cds} & Remark \\ \hline
        $\mathbb{PG}(2,2)$ & 7 & $3/7$ & $\{0,1,3\}$ & $m$-sequence, Barker code \\ \hline
        $\mathbb{PG}(2,3)$ & 13 & $4/13$ & $\{0,1,4,6\}$ & Barker code \\ \hline 
        $\mathbb{PG}(3,2)$ & 15 & $7/15$ & $\{0, 1, 2, 7, 9,12,13\}$ & $m$-sequence \\ \hline
        $\mathbb{PG}(3,3)$ & 40 & $13/40$ & $\{0, 1, 2, 4, 5, 8,13,14,17,19,24,26,34\}$ & \\\hline
    \end{tabular}
    \vspace{-5mm}
\end{table*}
Proposition \ref{prop:eavsl} suggests that expected \ac{isl} is not an appropriate sidelobe performance metric for designing the transmission mask $\V{m}_{\rm t}$. In this subsection, we investigate the \ac{pesl} characteristics of \ac{masm} signals. The transmission mask design problem minimizing the \ac{pesl} can be formulated using terminologies similar to those in the \ac{cds} formalism, as follows:
\begin{subequations}\label{pesl_min_ori}
\begin{align}
\min_{\Set{C}} & \max_{k>0,l>0,l\neq k} \left|(k\oplus\Set{C})\cap (l\oplus\Set{C})\right|-\left|(k\oplus\Set{C})\cap (l\oplus\Set{C})\cap \Set{C}\right|,\\
{\rm s.t.}&~~|\Set{C}|=\rho N,~\Set{C}\subseteq \mathbb{Z}_N,
\end{align} 
\end{subequations}
with the transmission mask being the indicator function of the set $\Set{C}$, i.e., $m_{{\rm t},i}=\mathbb{I}\{i+1\in \Set{C}\}$. For mainlobe-fluctuation-ideal transmission masks, the term $\left|(k\oplus\Set{C})\cap (l\oplus\Set{C})\right|$ is a constant according to the definition of \acp{cds}, and hence the \ac{pesl} minimisation problem can be simplified as
\begin{subequations}\label{pesl_min_simple}
\begin{align}
\max_{\Set{C}} & \min_{k>0,l>0,l\neq k} \left|(k\oplus\Set{C})\cap (l\oplus\Set{C})\cap \Set{C}\right|,\\
{\rm s.t.}&~~|\Set{C}|=\rho N,~\Set{C}\subseteq \mathbb{Z}_N.
\end{align} 
\end{subequations}

In general, the problems \eqref{pesl_min_ori} and \eqref{pesl_min_simple} do not admit efficient and exact solutions except for small-$N$ cases. Similar to the discussion of mainlobe fluctuation minimisation problem, one may then be interested in the existence and constructions of \ac{pesl}-ideal, or even simultaneously \ac{pesl}-ideal and mainlobe-fluctuation-ideal, transmission masks, under certain configurations of $(N,\rho)$. To this end, we should first concretise the concept of \ac{pesl}-ideal transmission masks by finding an achievable lower bound for \ac{pesl}. Using Proposition \ref{prop:eavsl}, the following lower bound on the \ac{pesl} can be obtained straightforwardly.
\begin{corollary}[Ideal \ac{pesl}]\label{coro:lb_pesl}
For any transmission mask $\V{m}_{\rm t}\in\{0,1\}^N$ satisfying $\V{1}^{\rm T}\V{m}_{\rm t}=\rho N$, the \ac{pesl} is lower bounded by
\begin{equation}\label{lb_pesl}
\max_{k>0,l>0,l\neq k} \mathbb{E}\{|r_{k,l}|^2\}\geq \left\lceil\frac{\rho(1-\rho)N^2(\rho N-1)}{(N-1)(N-2)}\right\rceil.
\end{equation}
\begin{IEEEproof}
Apparently, \ac{pesl} is no smaller than the \ac{aesl}, namely we have
$$
\max_{k>0,l>0,l\neq k} \mathbb{E}\{|r_{k,l}|^2\}\geq \frac{\rho(1-\rho)N^2(\rho N-1)}{(N-1)(N-2)}.
$$
Now note that all entries $[\M{R}]_{k,l}$ are integers, thereby the result \eqref{lb_pesl} follows from \eqref{esl}.
\end{IEEEproof}
\end{corollary}
Next we show that the lower bound \eqref{lb_pesl} is achievable (and hence is tight) for certain pairs of $(N,\rho)$, by means of explicit construction. Moreover, these transmission masks are simultaneously \ac{pesl}-ideal and mainlobe-fluctuation-ideal. 

\begin{proposition}[Singer \acp{cds} are ideal transmission masks]\label{prop:singer_ideal}
The transmission masks obtained by Singer's construction of \acp{cds} \cite{singer1938theorem} are simultaneously mainlobe-fluctuation-ideal and \ac{pesl}-ideal. To elaborate, a transmission mask $\V{m}_{\rm t}$, whose corresponding convolution operator given by
\begin{equation}\label{convolution_pg}
\M{A} = \M{F}{\rm diag}(\M{F}\V{m}_{\rm t})\M{F}^{\rm H} = [\V{m}_{\rm t};\widetilde{\M{J}}_1\V{m}_{\rm t};\dotsc;\widetilde{\M{J}}_{N-1}\V{m}_{\rm t}]
\end{equation}
is one of the possible incidence matrices between the points and hyperplanes (i.e., $(n-1)$-dimensional subspaces) in the finite projective space $\mathbb{PG}(n,q)$ (where $q$ is a power of prime) with $N=(q^{n+1}-1)(q-1)^{-1}$, is both mainlobe-fluctuation-ideal and \ac{pesl}-ideal. The duty cycles of such transmission masks are determined by $n$ and $q$ according to
\begin{equation}\label{duty_cycle_singer}
\rho = \frac{q^n-1}{N(q-1)}.
\end{equation}
\begin{IEEEproof}
Please refer to Appendix \ref{sec:proof_singer_ideal}.
\end{IEEEproof}
\end{proposition}

Proposition \ref{prop:singer_ideal} indicates that whenever the corresponding $(N,\rho)$ pairs are acceptable, the Singer \acp{cds} should be chosen as the transmission masks. The family of Singer \acp{cds} incorporates all $m$-sequences and some Barker codes. A subset of Singer \acp{cds} is summarised in Table \ref{tab:singer_examples}. Readers interested in a detailed charcterisation of these \acp{cds} are referred to \cite{dillon2004new}.

\subsection{Comparison with the Full-Duplex Scheme}\label{ssec:fd_comparison}
Now that we have characterised the range response (including both mainlobes and sidelobes) of half-duplex transmission masks, we may compare their performance with full-duplex transmission schemes. According to \cite{FanLiu2024}, the expected periodic \ac{acf} sidelobe level of full-duplex transmission schemes is simply given by\footnote{In fact, \eqref{fd_sc} is the expression for single-carrier full-duplex transmission schemes. We quote this formula for the fairness of comparison, since half-duplex schemes cannot easily apply other modulation bases due to their non-continuous temporal support.}
\begin{equation}\label{fd_sc}
\mathbb{E}\{|r_{k,l}|^2\} = N,~\forall k\neq l.
\end{equation}
Their corresponding expected mainlobe level is
$$
\mathbb{E}\{|r_{k,k}|^2\} = N^2+(\mu_4-1)N,~\forall k.
$$
Thus the mainlobe-to-\ac{pesl} ratio and the mainlobe-to-\ac{aesl} ratio depicting target detection dynamic range are equal to each other for full-duplex schemes, given by
\begin{equation}
\frac{\mathbb{E}\{|r_{k,k}|^2\}}{\mathbb{E}\{|r_{k,l}|^2\}} = N+\mu_4-1,~\forall k\neq l.
\end{equation}
By contrast, for half-duplex schemes, \ac{aesl} and \ac{pesl} are not necessarily equal. Furthermore, the mainlobe-to-\ac{pesl} ratio and the mainlobe-to-\ac{aesl} ratio are only appropriately applicable to mainlobe-fluctuation-ideal transmission masks. In particular, the mainlobe-to-\ac{aesl} ratio of mainlobe-fluctuation-ideal transmission masks is given by
\begin{align}\label{atmr}
\frac{\mathbb{E}\{|r_{k,k}|^2\}}{\mathbb{E}\{|r_{k,l}|^2\}}&=\frac{\rho (1-\rho)(N-1)(N-2)}{\rho N-1} \nonumber\\
&\hspace{3mm}+ (\mu_4\!-\!1)\cdot\frac{N\!-\!2}{N\!-\!1},~\forall k\neq l,~k\!>\!0,~l\!>\!0.
\end{align}
As $N$ tends to infinity, it follows that
\begin{equation}\label{asymptotic_atmr}
\frac{\mathbb{E}\{|r_{k,k}|^2\}}{\mathbb{E}\{|r_{k,l}|^2\}} \sim (1-\rho)N + \mu_4+1,~\forall k\neq l,~k>0,~l>0.
\end{equation}
The mainlobe-to-\ac{pesl} ratio is in general smaller, but \ac{pesl}-ideal transmission masks can achieve the upper bound given by \eqref{atmr}, thereby the asymptotic behaviour \eqref{asymptotic_atmr} also applies. 
\begin{remark}[Throughput-Sidelobe-Inteference Tradeoff]
The asymptotic behaviour \eqref{asymptotic_atmr} suggests a communication-sensing performance tradeoff: The communication throughput is proportional to $\rho$, but the target detection dynamic range is asymptotically proportional to $(1-\rho)$. As a typical operating point, the $m$-sequences have roughly $\rho=1/2$ for large $N$, which achieves a $50\%$ communication throughput and a 3dB loss of sensing dynamic range compared to the full-duplex scheme. These performance degradations can be viewed as the cost paid to obtain \ac{si}-free long-range sensing capability.
\end{remark}

\section{\ac{masm} Applied in Slow Time}\label{sec:slow_time}
In the aforementioned discussions, we have considered the transmission mask design in the fast time. Unfortunately, practical hardware may not be capable of supporting a sample-level switching rate between transmitting and receiving modes. To this end, in this section, we consider the slow-time transmission mask design problem. In particular, we confine ourselves to the transmission masks taking the following piecewise constant form:
\begin{equation}\label{piecewise_constant}
\V{m}_{\rm t} = \widetilde{\V{m}}_{\rm t} \otimes \V{1}_{T},
\end{equation}
where $\widetilde{\V{m}}_{\rm t}\in\{0,1\}^L$ represents the slow-time transmission mask, while $T$ denotes the number of fast-time samples contained in each slow-time symbol, satisfying $LT=N$. Similar to $\V{m}_{\rm t}$, we denote the $i$-th entry of $\widetilde{\V{m}}_{\rm t}$ as $\widetilde{m}_{{\rm t},i}$.

\subsection{Range Glint Analysis}
In the fast-time version of \ac{masm}, we defined the expected \ac{irgi} as a metric of mainlobe fluctuation, which characterises the mainlobe variation outside the inevitable blind range $k=0$. Due to the piecewise constant structure \eqref{piecewise_constant}, slow-time coding-based transmission strategies would yield blind ranges approximately as large as $T$, namely the length of a sub-pulse, corresponding to the ``ramping up'' region in Fig.~\ref{fig:trapezoid}. In light of this, we define the expected partial \ac{rgi} as
\begin{equation}
{\rm EPRGI}_k(\Set{B}) = \mathbb{E}\bigg\{\Big|r_{k,k}-\frac{1}{N-|\Set{B}|}\sum_{l\in\overline{\Set{B}}}r_{l,l}\Big|^2\bigg\},
\end{equation}
where $\Set{B}$ and $\overline{\Set{B}}$ denote the sample indices corresponding to the blind range and the complement of the blind range (relative to $\{1,2,\dotsc,N\}$), respectively. Before analysing the mainlobe fluctuation performance of the slow-time coding strategy, we first present a simplified expression for the expected partial \ac{rgi}.
\begin{lemma}\label{lem:eprgi}
The expected partial \ac{rgi} can be expressed as
\begin{align}\label{eprgi_full}
{\rm EPRGI}_k(\Set{B}) &= \bigg(a_k-\frac{1}{N-|\Set{B}|}\sum_{l\in\overline{\Set{B}}}a_l\bigg)^2+ (\mu_4-1) \nonumber\\
&\hspace{3mm}\times \bigg[-a_k+\frac{2}{N-|\Set{B}|}\sum_{q\in\overline{\Set{B}}}b_{kq}+\frac{1}{(N-|\Set{B}|)^2} \nonumber \\
&\hspace{7mm}\times \sum_{p^\prime\in\overline{\Set{B}}}\sum_{q^\prime\in\overline{\Set{B}}}(a_{p^\prime}-b_{p^\prime q^\prime})\bigg],
\end{align}
where $a_k = \V{1}^{\rm T}[(\V{1}-\V{m}_{\rm t})\odot \widetilde{\M{J}}_k\V{m}_{\rm t}]$ denotes the cross-correlation between transmission and reception masks, and $b_{lp} = \V{1}^{\rm T}[(\V{1}-\V{m}_{\rm t})\odot \widetilde{\M{J}}_l\V{m}_{\rm t}\odot \widetilde{\M{J}}_{l-p}\V{m}_{\rm t}]$.
\begin{IEEEproof}
Please refer to Appendix \ref{sec:proof_eprgi}.
\end{IEEEproof}
\end{lemma}

From Lemma \ref{lem:eprgi} we may observe that, the first term on the right-hand side of \eqref{eprgi_full} is on the order of $O(N^2)$, while the second term (containing the factor $(\mu_4-1)$) is on the order of $O(N)$. This resembles the decomposition discussed in Sec.~\ref{ssec:solving_rgm}. We may then conclude that, the key to reducing the expected partial \ac{rgi} is to suppress the fluctuation between $a_k$'s. 
\begin{remark}
In particular, we may say that a slow-time transmission mask $\widetilde{\V{m}}_{\rm t}$ is mainlobe-fluctuation-ideal up to the first order, whenever it satisfies
\begin{equation}\label{rgi_1st}
a_k=\frac{1}{N-|\Set{B}|}\sum_{l\in\overline{\Set{B}}}a_l,~\forall k\in\overline{\Set{B}}.
\end{equation}
For these transmission masks, the mainlobe fluctuation outside the blind range can be exactly zero under constant modulus constellations, namely $\mu_4=1$.
\end{remark} 

Next, we give a simple characterisation of $\{a_i\}_{i=0}^{N-1}$ under the piecewise constant constraint \eqref{piecewise_constant}.
\begin{proposition}\label{prop:mask_ccs}
The mask cross-correlation sequence $\{a_i\}_{i=0}^{N-1}$ satisfies
\begin{equation}\label{mask_ccs}
a_{kT+l} = T\widetilde{a}_k + l \cdot d_k,
\end{equation}
where $k=0,\dotsc,L-1$, $l=0,\dotsc,T-1$, $\widetilde{a}_k$ is the slow-time mask cross-correlation sequence defined as
$\widetilde{a}_k = \V{1}^{\rm T}[(\V{1}-\widetilde{\V{m}}_{\rm t})\odot \widetilde{\M{J}}_k\widetilde{\V{m}}_{\rm t}]$, while $d_k$ denotes the cyclic difference of $\widetilde{a}_k$ given by
$$
d_k = \widetilde{a}_{(k+1){\rm mod}N}- \widetilde{a}_k.
$$
\begin{IEEEproof}
Please refer to Appendix \ref{sec:proof_mask_ccs}.
\end{IEEEproof}
\end{proposition}

Since $\widetilde{a}_0=0$ holds for all slow-time transmission masks, it then follows from Proposition \ref{prop:mask_ccs} that when $k=0$, or $k=L$ and $l>0$, $a_{kT+l}$ would suffer from inevitable mainlobe fluctuation. Therefore, we define the blind range of such transmission masks as
\begin{equation}\label{blind_zone}
\Set{B} = \{0,\dotsc,T-1\}\cup \{(L-1)T+1,\dotsc, LT-1\}.
\end{equation}
Observe that \eqref{blind_zone} coincide with the blind range of conventional pulse radars with pulse width $T$. However, such systems can only transmit $T$ communication symbols in each \ac{pri}. By contrast, the proposed slow-time coding scheme significantly improves the communication throughput to $\rho LT$ symbols per \ac{pri}.

Another important insight provided by Proposition \ref{prop:mask_ccs} is that, in order to guarantee the first-order mainlobe fluctuation ideality depicted by \eqref{rgi_1st}, it suffices to ensure that
\begin{equation}
\widetilde{a}_k=\frac{1}{L-1}\sum_{l=1}^{L-1}\widetilde{a}_l,~\forall k=1,\dotsc,L-1.
\end{equation}
In other words, if $\widetilde{\V{m}}_{\rm t}$ is mainlobe-fluctuation-ideal when applied as an ordinary transmission mask, it is also mainlobe-fluctuation-ideal up to the first order when used as a slow-time transmission mask. Therefore, our discussions and constructions of ideal transmission masks in the previous sections can also be applied to slow-time coding schemes.

\subsection{Sidelobe Analysis}
Compared to the mainlobe fluctuation analysis, the sidelobe analysis of slow-time coding schemes is much simpler. In particular, the expression \eqref{esl} for the expected sidelobe level is still applicable. The remaining task is to further simplify the expression of $\M{R}$ via exploiting the piecewise constant structure of $\V{m}_{\rm t}$. In fact, by applying similar arguments as those used in Proposition \ref{prop:mask_ccs}, we may obtain the following result. 
\begin{corollary}\label{coro:sidelobe2d_st}
Define the slow-time sidelobe level as
$$
\widetilde{a}_{k,l} = \sum_{n=1}^N \widetilde{m}_{{\rm t},\overline{n-l}} \widetilde{m}_{{\rm t},\overline{n-k}} (1-\widetilde{m}_{{\rm t},n}).
$$
The expected sidelobe level under the slow-time coding scheme can then be written as
\begin{equation}\label{st_esl}
\mathbb{E}\{|r_{k_1T+l_1,k_2T+l_2}|^2\} = T\widetilde{a}_{k_1,k_2} + l_1\cdot d^{(1)}_{k_1,k_2}+ l_2\cdot d^{(2)}_{k_1,k_2},
\end{equation}
for $k_1T+l_1\neq 0$ and $k_2T+l_2\neq 0$, where
$$
\begin{aligned}
d_{k_1,k_2}^{(1)} &= \widetilde{a}_{(k_1+1){\rm mod}N,k_2}- \widetilde{a}_{k_1,k_2}, \\
d_{k_1,k_2}^{(2)} &= \widetilde{a}_{k_1,(k_2+1){\rm mod}N}- \widetilde{a}_{k_1,k_2}.
\end{aligned}
$$
\begin{IEEEproof}
This result can be obtained by applying again the arguments used for proving Proposition \ref{prop:mask_ccs}, detailed in Appendix \ref{sec:proof_mask_ccs}.
\end{IEEEproof}
\end{corollary}

Corollary \ref{coro:sidelobe2d_st} implies that, the aforementioned discussions in Sec.~\ref{sec:sidelobe} concerning average sidelobe levels and peak sidelobe levels can also be applied to slow-time coding schemes via \eqref{st_esl}. In light of this, Singer \acp{cds} are also ideal slow-time transmission masks in the sense detailed in Proposition \ref{prop:singer_ideal}.

\section{Numerical Results}\label{sec:numerical}
In this section, we verify and demonstrate the analytical results discussed in previous sections using numerical examples.
\subsection{Fast-Time \ac{masm}}
\begin{figure}[t]
    \centering
    \includegraphics[width=0.85\linewidth]{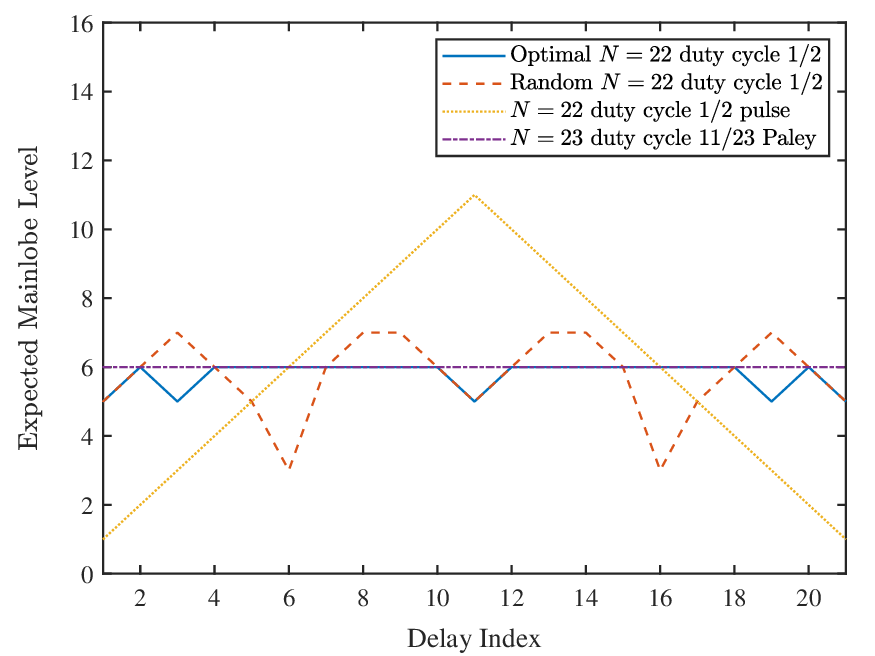}
    \caption{The mainlobe fluctuation of \ac{masm} under PSK constellations and different transmission masks, compared with the conventional pulse.}
    \label{fig:range_glint_fasttime}
    \vspace{-4mm}
\end{figure}

Let us first consider the mainlobe fluctuation effect. In Fig.~\ref{fig:range_glint_fasttime}, the expected mainlobe levels (i.e., $\mathbb{E}\{|r_{k,k}|^2\}$) of \ac{masm} having different transmission masks, as well as that of the conventional pulse, are portrayed. We consider PSK constellations, thereby the expected mainlobe level is sufficient to characterise mainlobe fluctuation (see Proposition \ref{prop:eirgi}). Observe that the mainlobe fluctuation of the conventional pulse and \ac{masm} with random transmission mask is rather significant, which is unsatisfactory. The optimal $N=22$, $\rho=1/2$ transmission mask exhibits much weaker mainlobe fluctuation, with only $5$ delay bins deviating from the most likely mainlobe level. The configuration $N=23$, $\rho=11/23$ admits a mainlobe-fluctuation-ideal transmission mask (the Paley \ac{cds}) whose expected mainlobe level is constant across all delay bins $0<k\leq N-1$.

\begin{figure}[t]
    \centering
    \includegraphics[width=0.85\linewidth]{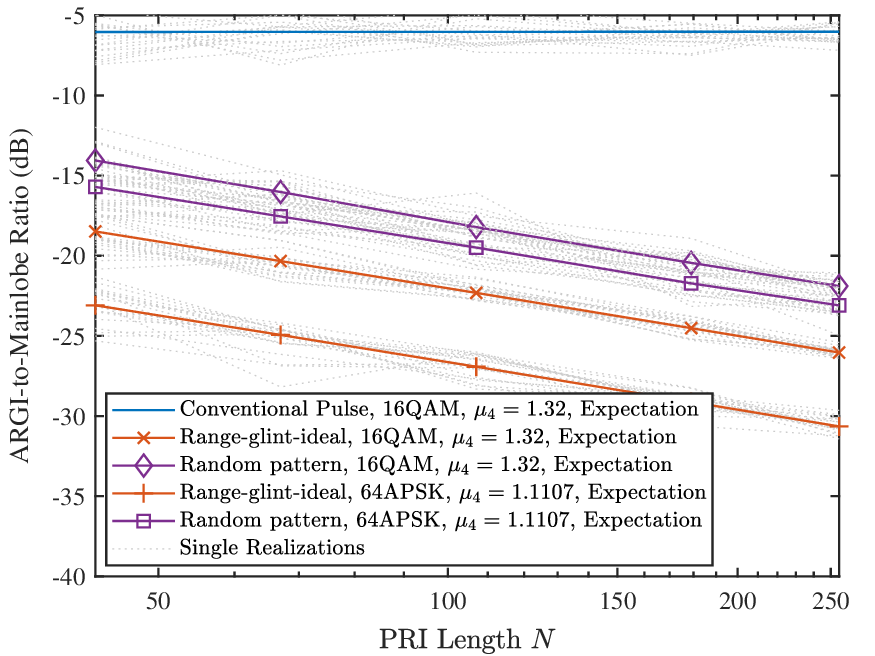}
    \caption{The \ac{irgi}-to-mainlobe ratio of \ac{masm} under various constellations and transmission masks, compared with the conventional pulse.}
    \label{fig:eargi_all}
    \vspace{-4mm}
\end{figure}

To demonstrate the scaling laws given in Table~\ref{tab:eirgi_scaling}, the \ac{irgi}-to-mainlobe ratios of \ac{masm} under various constellations and transmission masks are compared against that of the conventional pulse in Fig.~\ref{fig:eargi_all}. The ratio is defined as $N\cdot {\rm ARGI}\cdot (\sum_{k=0}^{N-1}\mathbb{E}\{|r_{k,k}|^2\})^{-1}$. To ensure the fairness of comparison, all transmission masks have duty cycle approximately equal to $0.5$. Observe that both random transmission masks and mainlobe-fluctuation-ideal masks exhibit $O(1/N)$ \ac{irgi}-to-mainlobe ratios, while the conventional pulses have constant \ac{irgi}-to-mainlobe ratios. Furthermore, the ratio reduces as $\mu_4$ decreases, as implied by Proposition \ref{prop:eirgi}.

\begin{figure}[t]
\vspace{-4mm}
\subfloat[][Singer $\mathbb{PG}(5,2)$.]{
\centering
\includegraphics[width=.48\columnwidth]{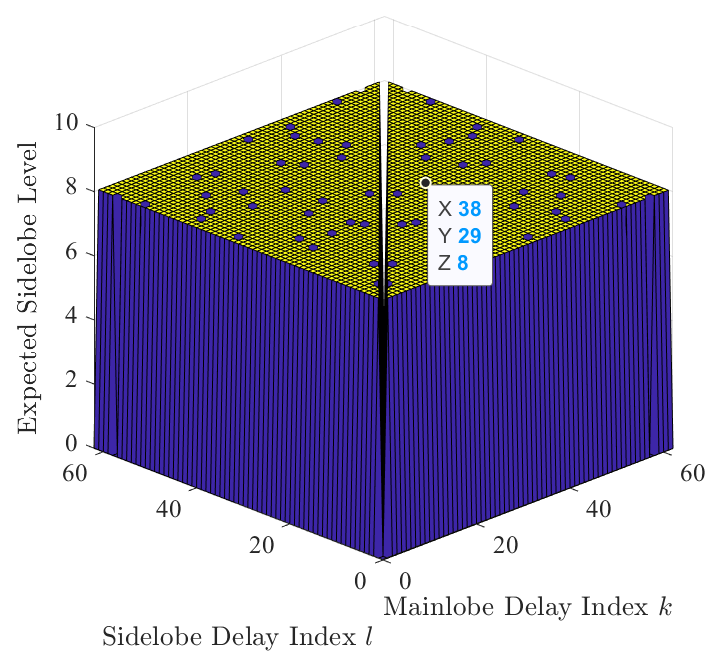}
\label{fig:singerR}
}
\subfloat[][GMW Sequence.]{
\centering
\includegraphics[width=.48\columnwidth]{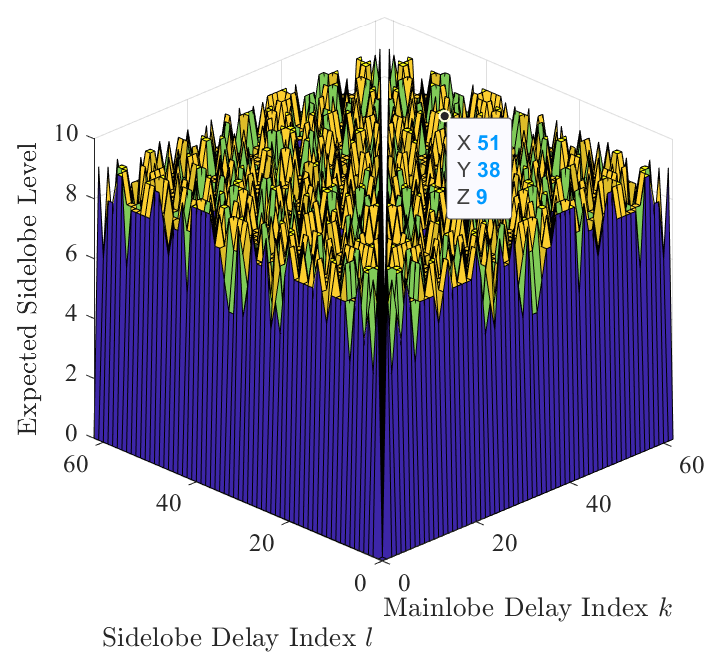}
\label{fig:gmwR}
}
\caption{The expected sidelobe levels of two different mainlobe-fluctuation-ideal transmission masks, for which the $(N,\rho)$ configuration is $N=63$ and $\rho=31/63$.}
\label{fig:two_sidelobe_levels}
\vspace{-4mm}
\end{figure}

Next, we investigate the sidelobe performance. In Fig.~\ref{fig:two_sidelobe_levels}, the expected sidelobe levels of two different mainlobe-fluctuation-ideal $N=63$, $\rho=31/63$ transmission masks are portrayed. According to Proposition \ref{prop:eavsl}, the \ac{aesl} is identical for both transmission masks, which equals $\frac{29760}{3782}\approx7.869$. However, the Singer $\mathbb{PG}(5,2)$ sequence in Fig.~\ref{fig:singerR} is \ac{pesl}-ideal, whose expected sidelobe level is at most $\lceil 7.869\rceil = 8$. By contrast, the GMW sequence in Fig.~\ref{fig:gmwR} has a \ac{pesl} of $9$.

\begin{figure}[t]
    \centering
    \includegraphics[width=0.85\linewidth]{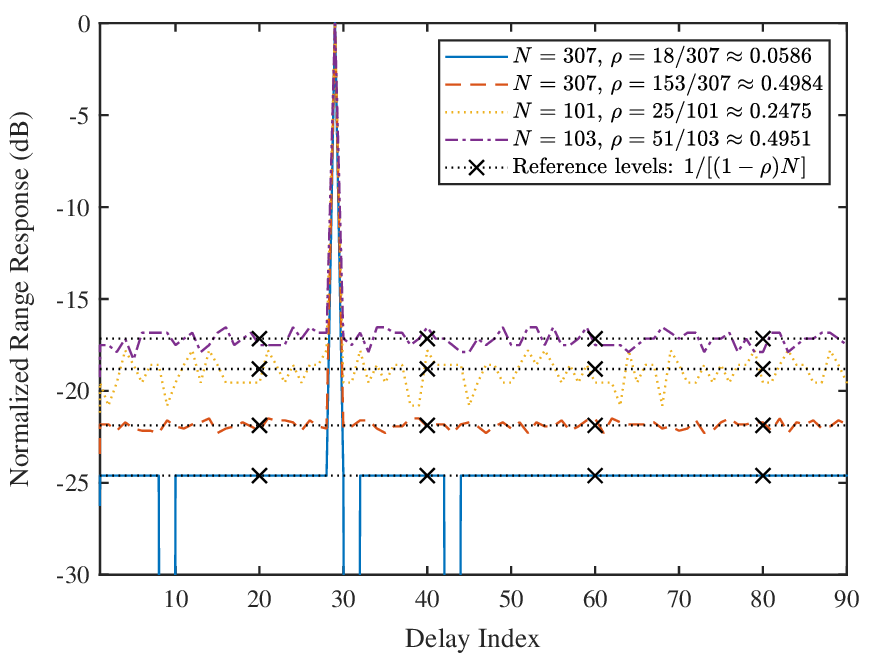}
    \caption{Range response slices of \ac{masm} under different $(N,\rho)$ configurations.}
    \label{fig:sidelobe_slices_fasttime}
    \vspace{-4mm}
\end{figure}

Finally, let us verify and demonstrate the results in Sec.~\ref{ssec:fd_comparison}. In particular, we choose several mainlobe-fluctuation-ideal transmission masks with diverse $(N,\rho)$ configurations, and plot their range response slices at mainlobe delay $k=29$ in Fig.~\ref{fig:sidelobe_slices_fasttime}. The constellations are chosen to be QPSK. It is seen that the expected sidelobe levels can be fit well to the asymptotic reference sidelobe level of $1/[(1-\rho)N]$, as depicted by \eqref{asymptotic_atmr}.

\subsection{Slow-Time \ac{masm}}
\begin{table}[t]
\caption{The parameters used for \ac{prf} staggering methods.}
    \label{tab:staggeredPRI}
    \centering
    \begin{tabular}{|c|c|c|c|} \hline
        Stagger ratio & \makecell{No. of periods \\ for each \ac{prf}} & Pulse width & Duty cycle \\ \hline
        $31:32:33$ & 11 & 16 & 0.500 \\ \hline
        $25:30:27:31$ & 11 & 14 & 0.496 \\ \hline
        $51:62:53:61:58$ & 4 & 28 & 0.491 \\ \hline
    \end{tabular}
    \vspace{-4mm}
\end{table}
In this subsection, we investigate the performance of \ac{masm} as a slow-time coding technique. The benchmark is the \ac{prf} staggering methods, with stagger ratios of $31:32:33$, $25:30:27:31$ and $51:62:53:61:58$, adopted from \cite[Sec.~15.9]{radar_handbook}. The transmission mask for \ac{masm} is chosen to be the 63-point $m$-sequence, whose duty cycle is $31/63\approx 0.492$. Each slow-time sample contains $T=16$ symbols, resulting in $N=1008$. To ensure that the total sequence lengths as well as the duty cycles of the considered methods are approximately equal to one another, we use the parameters in Table~\ref{tab:staggeredPRI} for the \ac{prf} staggering methods.

\begin{figure}[t]
    \centering
    \includegraphics[width=0.85\linewidth]{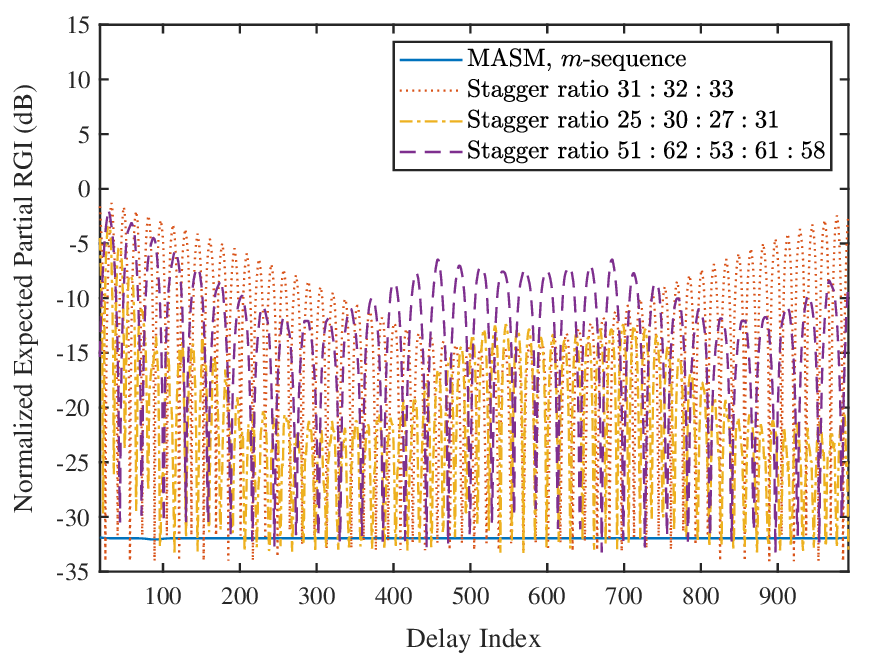}
    \caption{The mainlobe fluctuation of slow-time \ac{masm} compared with \ac{prf} staggering methods, under 16QAM constellations, measured by normalised expected partial \ac{rgi}.}
    \label{fig:eprgi}
    \vspace{-4mm}
\end{figure}

We first consider the mainlobe fluctuation effect. In Fig.~\ref{fig:eprgi}, the slow-time \ac{masm} is compared with \ac{prf} staggering methods, in terms of the expected partial \ac{rgi}, which is normalised in the following manner
\begin{equation}
{\rm NEPRGI}_k(\Set{B}) = \frac{\mathbb{E}\bigg\{\Big|r_{k,k}-\frac{1}{N-|\Set{B}|}\sum_{l\in\overline{\Set{B}}}r_{l,l}\Big|^2\bigg\}}{\mathbb{E}\{|r_{k,k}|^2\}}.
\end{equation}
In this figure, we consider the 16QAM constellation. Observe that the mainlobe fluctuation effect of \ac{masm} is much weaker than that of \ac{prf} staggering methods. This can be understood by revisiting \eqref{eprgi_full}, in which the first term on the right-hand side is zero for \ac{masm}. Therefore, the expected partial \ac{rgi} of \ac{masm} is $O(N)$ times lower than that of \ac{prf} staggering methods. 

\begin{figure}[t]
    \centering
    \includegraphics[width=0.85\linewidth]{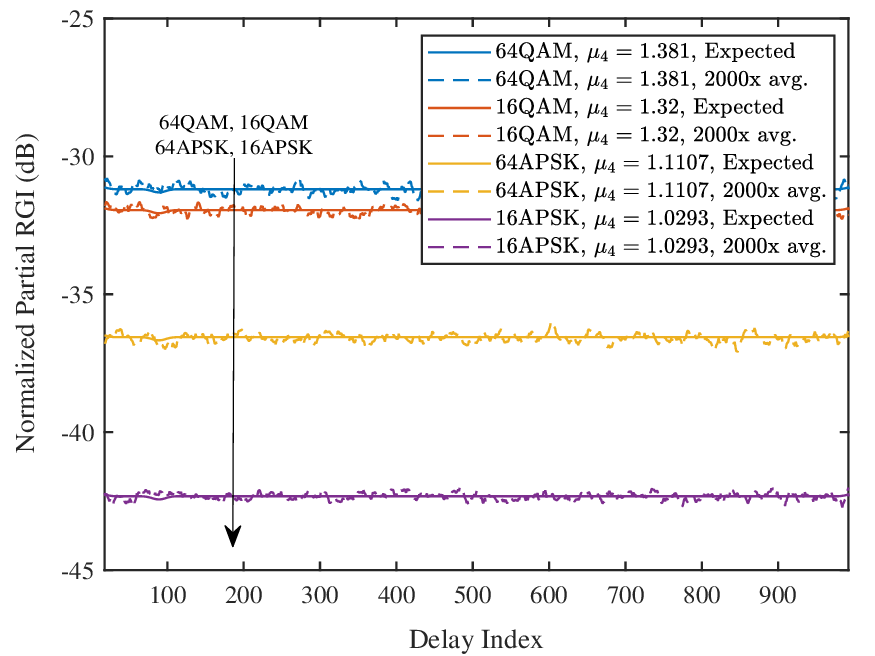}
    \caption{The averaged and expected partial \ac{rgi} of \ac{masm} under different constellations.}
    \label{fig:prgi_stpm}
    \vspace{-4mm}
\end{figure}

Next, let us investigate the relationship between mainlobe fluctuation and constellations, as portrayed in Fig.~\ref{fig:prgi_stpm}. Note that for \ac{masm}, the expected partial \ac{rgi} is proportional to $(\mu_4-1)$. As seen in Fig.~\ref{fig:prgi_stpm}, the normalised expected partial \ac{rgi} indeed decreases with the fourth moment $\mu_4$ of the constellations. Moreover, the partial \ac{rgi} averaged over 2000 random instances are in close proximity to their corresponding analytical expectations.

\begin{figure}[t]
    \centering
    \includegraphics[width=0.85\linewidth]{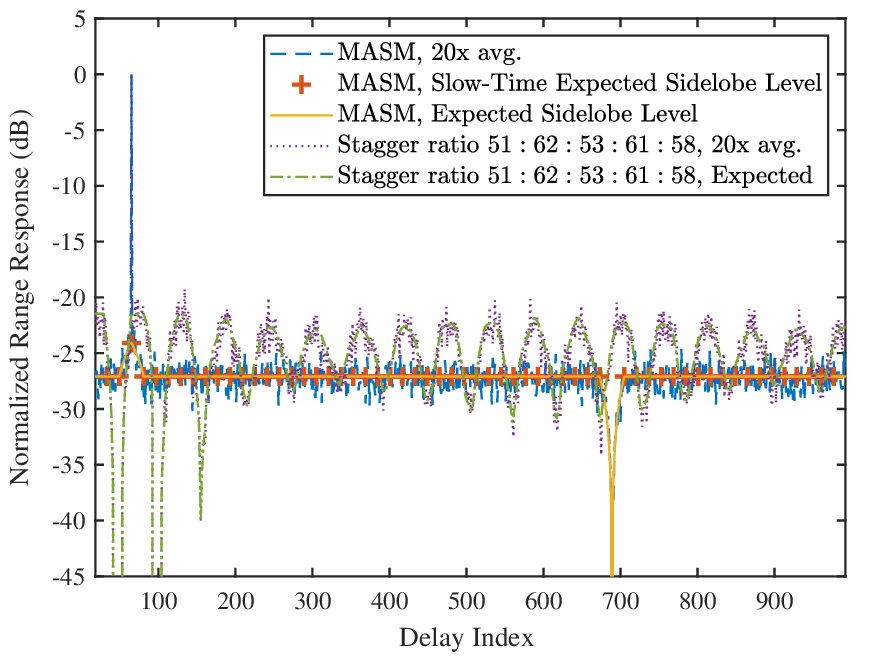}
    \caption{The averaged and expected sidelobe level of \ac{masm} compared with that of \ac{prf} staggering methods.}
    \label{fig:sidelobe_slowtime}
    \vspace{-4mm}
\end{figure}

Finally, we consider the sidelobe performance of slow-time \ac{masm}. In Fig.~\ref{fig:sidelobe_slowtime}, \ac{masm} is compared against the \ac{prf} staggering method having stagger ratio $51:62:63:61:58$, in terms of sidelobe level. The mainlobe is located at the $65$-th delay bin. It is clear that the \ac{pesl} of slow-time \ac{masm} is significantly lower than that of the \ac{prf} staggering method. Furthermore, the sidelobe levels at integer multiples of the slow-time sample period coincide with the (upsampled) sidelobe levels of the slow-time transmission mask, which corroborates Corollary \ref{coro:sidelobe2d_st}.

\section{Conclusions}\label{sec:conclusion}
In this treatise, we have proposed \ac{masm}, a half-duplex \ac{isac} transmission scheme avoiding the \ac{si} issue in long-range sensing. Compared to conventional pulse radar techniques and the \ac{prf} staggering techniques, \ac{masm} can mitigate the mainlobe fluctuation effect, which characterises the range blindness, to a significantly lower level, while preserving large duty cycles ($\sim$50\%). This empowers \ac{isac} systems with high communication throughput and high energy accumulation efficiency. Furthermore, both mainlobe-fluctuation-ideal and sidelobe-ideal transmission masks are constructed, which can be readily used in practical \ac{isac} systems. Our hope is that \ac{masm} may be beneficially applied in future long-range \ac{isac} scenarios, including low-altitude economy and autonomous driving.

\appendices
\section{Range Glint Analysis}
\subsection{Proof of Proposition \ref{prop:det_irgi}}\label{sec:proof_det_irgi}
\begin{IEEEproof}
The \ac{irgi} can be expressed as
\begin{align}
\overline{g}(\V{r}_{\rm m})&= \frac{1}{N-1}\sum_{k=1}^{N-1} |r_{k,k}|^2 - \left|\frac{1}{N-1} \sum_{l=1}^{N-1} r_{l,l}\right|^2\nonumber\\
%&\hspace{2mm}=\frac{1}{N-1} (\|\V{r}_{\rm m}\|_2^2 - |r_{0,0}|^2)- \frac{1}{(N-1)^2} \left|\V{1}^{\rm T}\V{r}_{\rm m}- r_{0,0}\right|^2\nonumber\\
&\hspace{2mm}= \frac{N}{N-1}\left[\|\M{F}\V{m}_{\rm t}\|_4^4 -\frac{1}{N}(\V{1}^{\rm T}|\M{F}\V{m}_{\rm t}|^2)^2\right] \nonumber\\
&\hspace{5mm}- \frac{N^2}{(N-1)^2} \left(\rho^2N-\frac{1}{N}\V{1}^{\rm T}|\M{F}\V{m}_{\rm t}|^2\right)^2.
\end{align}
Note that $\rho^2 N$ is constant with respect to $\V{m}_{\rm t}$. Using $\V{1}^{\rm T}|\M{F}\V{m}_{\rm t}|^2=\|\V{m}_{\rm t}\|_2^2$, we may rearrange the expression of $\overline{g}(\V{r}_{\rm m})$ by separating the terms containing $\V{m}_{\rm t}$ from the constants, as follows:
\begin{align}\label{irgi_cm}
\overline{g}(\V{r}_{\rm m}) &= \frac{N}{N-1}\|\M{F}\V{m}_{\rm t}\|_4^4-\frac{1}{N-1}\|\V{m}_{\rm t}\|_2^4-\frac{\rho^4 N^4}{(N-1)^2}\nonumber\\
&\hspace{3mm}+\frac{2\rho^2 N^2}{(N-1)^2}\|\V{m}_{\rm t}\|_2^2-\frac{1}{(N-1)^2}\|\V{m}_{\rm t}\|_2^4\nonumber\\
&=\frac{N}{N-1}\|\M{F}\V{m}_{\rm t}\|_4^4-\frac{\rho^2 N^3\left(\rho^2N-2\rho+1\right)}{(N-1)^2},
\end{align}
which completes the proof.
\end{IEEEproof}

\subsection{Proof of Proposition \ref{prop:eirgi}}\label{sec:proof_eirgi}
\begin{IEEEproof}
According to \eqref{mainlobe}, we have
$$
r_{k,k} = \left\|(\V{1}-\V{m}_{\rm t})\odot \left[\widetilde{\M{J}}_k(\V{m}_{\rm t}\odot\V{x}_i)\right]\right\|_2^2.
$$
Thereby the generic $\V{r}_{\rm m}$ for half-duplex \ac{isac} systems can be expressed as
\begin{align}
\V{r}_{\rm m} &= \sqrt{N}\M{F}^{\rm H}\left[(\M{F}|\V{m}_{\rm t}\odot\V{x}_i|^2)^\ast\odot \M{F}(\V{1}-\V{m}_{\rm t})\right]\nonumber\\
&=-\sqrt{N}\M{F}^{\rm H}\left[(\M{F}(\V{m}_{\rm t}\odot|\V{x}_i|^2))^\ast\odot \M{F}\V{m}_{\rm t}\right]\nonumber \\
&\hspace{3mm}+ \V{1}^{\rm T}(\V{m}_{\rm t}\odot |\V{x}_i|^2)\V{1}.
\end{align}
According to Proposition \ref{prop:irgi_property}, the \ac{irgi} can be alternatively expressed as $\overline{g}(\V{r}_{\rm m})=\overline{g}(\widetilde{\V{r}}_{\rm m})$, where
$$
\widetilde{\V{r}}_{\rm m} = \sqrt{N}\M{F}^{\rm H}\left[(\M{F}(\V{m}_{\rm t}\odot|\V{x}_i|^2))^\ast\odot \M{F}\V{m}_{\rm t}\right].
$$
This leads to a simpler form of the \ac{irgi}:
$$
\begin{aligned}
\overline{g}(\V{r}_{\rm m}) &= \frac{N}{N-1}\bigg[\|\M{F}^*(\V{m}_{\rm t}\odot|\V{x}_i|^2)^*\odot\M{F}\V{m}_{\rm t}\|_2^2\\
&\hspace{3mm}-\frac{1}{N}\left[\V{1}^{\rm T}\left((\M{F}^*(\V{m}_{\rm t}\odot|\V{x}_i|^2)^*)\odot \M{F}\V{m}_{\rm t}\right)\right]^2\bigg] \\
&\hspace{3mm}-\frac{N^2}{(N-1)^2} \bigg(\rho\V{1}^{\rm T}(\V{m}_{\rm t}\odot|\V{x}_i|^2)\\
&\hspace{3mm}-\frac{1}{N}\V{1}^{\rm T}\left((\M{F}^*(\V{m}_{\rm t}\odot|\V{x}_i|^2)^*)\odot \M{F}\V{m}_{\rm t}\right)\bigg)^2.
\end{aligned}
$$
The expectations of the terms constituting \ac{irgi} are as follows
\begin{align}
&\mathbb{E}\{\|\M{F}^*(\V{m}_{\rm t}\odot|\V{x}_i|^2)^*\odot\M{F}\V{m}_{\rm t}\|_2^2\}\nonumber\\
&\hspace{3mm}= \V{1}^{\rm T}\left(\mathbb{E}\left\{\left|\M{F}(\V{m}_{\rm t}\odot|\V{x}_i|^2)\right|^2\right\}\odot |\M{F}\V{m}_{\rm t}|^2\right)\nonumber\\
&\hspace{3mm}=\sum_{n=1}^N |\V{f}_n^{\rm T}\V{m}_{\rm t}|^2\V{f}_n^{\rm H}\M{M}_{\rm t}\M{S}\M{M}_{\rm t}\V{f}_n,
\end{align}
where $\M{S}:=\mathbb{E}\{|\V{x}_i|^2|\V{x}_i|^{2T}\}$, $\V{f}_n$ denotes the transpose of the $n$-th row of $\M{F}$,
\begin{equation}
\mathbb{E}\left\{[\V{1}^{\rm T}(\V{m}_{\rm t}\odot|\V{x}_i|^2)]^2\right\}=\V{m}_{\rm t}^{\rm T}\M{S}\V{m}_{\rm t}, \nonumber
\end{equation}
\begin{align}
\mathbb{E}\left\{\left[\V{1}^{\rm T}\left((\M{F}^*(\V{m}_{\rm t}\odot|\V{x}_i|^2)^*)\odot \M{F}\V{m}_{\rm t}\right)\right]^2\right\}=\V{m}_{\rm t}^{\rm T}\M{S}\V{m}_{\rm t}, \nonumber
\end{align}
and
\begin{align}
&\mathbb{E}\left\{\V{1}^{\rm T}(\V{m}_{\rm t}\odot|\V{x}_i|^2)\V{1}^{\rm T}\left((\M{F}^*(\V{m}_{\rm t}\odot|\V{x}_i|^2)^*)\odot \M{F}\V{m}_{\rm t}\right)\right\} \nonumber\\
&\hspace{3mm}=\V{m}_{\rm t}^{\rm T}\M{S}\V{m}_{\rm t}. \nonumber
\end{align}
Thus we have
\begin{align}
\mathbb{E}_{\V{x}_0}\{\overline{g}(\V{r}_{\rm m})\}&=\frac{N}{N-1}\sum_{n=1}^N|\V{f}_n^{\rm T}\V{m}_{\rm t}|^2\V{f}_n^{\rm H}\M{M}_{\rm t}\M{S}\M{M}_{\rm t}\V{f}_n\nonumber\\
&\hspace{3mm}-\frac{1}{N-1}\V{m}_{\rm t}^{\rm H}\M{S}\V{m}_{\rm t}-\frac{\rho^2N^2}{(N-1)^2}\V{m}_{\rm t}^{\rm H}\M{S}\V{m}_{\rm t} \nonumber\\
&\hspace{3mm}-\frac{1}{(N-1)^2}\V{m}_{\rm t}^{\rm H}\M{S}\V{m}_{\rm t}+\frac{2\rho N}{(N-1)^2}\V{m}_{\rm t}^{\rm H}\M{S}\V{m}_{\rm t}\nonumber\\
&=\frac{N}{N-1}\sum_{n=1}^N|\V{f}_n^{\rm T}\V{m}_{\rm t}|^2\V{f}_n^{\rm H}\M{M}_{\rm t}\M{S}\M{M}_{\rm t}\V{f}_n\nonumber \\
&\hspace{3mm}-\frac{N}{(N-1)^2}(\rho^2N-2\rho+1)\V{m}_{\rm t}^{\rm H}\M{S}\V{m}_{\rm t}.
\end{align}
We may now observe that the key to computing the expected \ac{irgi} is to obtain the matrix $\M{S}$, for which we have
\begin{equation}
\M{S}=\V{1}\V{1}^{\rm T} + (\mu_4-1)\M{I},
\end{equation}
where $\mu_4 = \mathbb{E}\{|x|^4\},~\forall x\in\Set{S}$. This implies \eqref{eirgi_data}.
\end{IEEEproof}

\subsection{Proof of Proposition \ref{prop:random_masks}}\label{sec:proof_random_masks}
\begin{IEEEproof}
It suffices to compute the expectation of \eqref{time_domain_eirgi}. Note that
$$
a_k a_l = \sum_{p=1}^N \sum_{q=1}^N (1-m_{{\rm t},p})(1-m_{{\rm t},q})m_{{\rm t},\overline{p-k}}m_{{\rm t},\overline{q-l}},
$$
where $\overline{n-l}$ denotes the cyclic shift relation of $(n-l-1)~{\rm mod}~N+1$. For each term in the above summation, the expectation can be computed as
\begin{equation}\label{complex_expectation}
\begin{aligned}
&\mathbb{E}\{(1-m_{{\rm t},p})(1-m_{{\rm t},q})m_{{\rm t},\overline{p-k}}m_{{\rm t},\overline{q-l}}\}\\
&\hspace{3mm}=\left\{
  \begin{array}{ll}
    \rho^2(1-\rho)^2, & \hbox{$p\neq q$, $k\neq l$, $\overline{p-k}\neq \overline{q-l}$;} \\
    \rho(1-\rho)^2, & \hbox{$p\neq q$, $k\neq l$, $\overline{p-k}= \overline{q-l}$;} \\
    \rho^2(1-\rho), & \hbox{$p=q$, $k\neq l$;} \\
    \rho^2(1-\rho)^2, & \hbox{$p\neq q$, $k=l$;} \\
    \rho(1-\rho), & \hbox{$p=q$, $k=l$.} \\
  \end{array}
\right.
\end{aligned}
\end{equation}
Using \eqref{complex_expectation}, and counting the terms of each type in the summation, we obtain
$$
\begin{aligned}
&\mathbb{E}\bigg\{\frac{1}{N-1}\sum_{k=1}^{N-1} \left(a_k-\frac{1}{N-1}\sum_{l=1}^{N-1}a_l\right)^2\bigg\}\\
&\hspace{2mm}=\frac{N-2}{N-1}(N(N-1)\rho^2(1-\rho)^2+N\rho(1-\rho)) \\
&\hspace{4mm}+\frac{1}{(N-1)^2}[ (N-1)(N-2)^2N\rho^2(1-\rho)^2 + (N-1)\\
&\hspace{6mm} \times (N-2)N\rho(1-\rho)^2 + (N-1)(N-2)N\rho^2(1-\rho)],
\end{aligned}
$$
which can be simplified into \eqref{eirgi_data2}.
\end{IEEEproof}

\subsection{Proof of Lemma \ref{lem:eprgi}}\label{sec:proof_eprgi}
\begin{IEEEproof}
Let us first write the expected partial \ac{rgi} as
\begin{align}\label{eprgi_expanded}
{\rm EPRGI}_k(\Set{B}) &= \mathbb{E}\{|r_{k,k}|^2\} -\frac{2}{N-|\Set{B}|}\sum_{q\in\overline{\Set{B}}}{\rm Re}\Big[\mathbb{E}\{(r_{k,k})^*r_{q,q}\}\Big] \nonumber\\
&\hspace{3mm}+ \frac{1}{(N-|\Set{B}|)^2}\sum_{l\in\overline{\Set{B}}}\sum_{p\in\overline{\Set{B}}}\mathbb{E}\{(r_{l,l})^*r_{p,p}\}.
\end{align}
The term $\mathbb{E}\{|r_{k,k}|^2\}$ can be written as
\begin{align}\label{self_terms}
\mathbb{E}\{|r_{k,k}|^2\}&= \mathbb{E}\bigg\{\left\|(\V{1}-\V{m}_{\rm t})\odot \left[\widetilde{\M{J}}_k(\V{m}_{\rm t}\odot\V{x}_i)\right]\right\|^2\bigg]\nonumber\\
&=\mathbb{E}\bigg[\V{1}^{\rm T}\Big[(\V{1}-\V{m}_{\rm t})\odot \widetilde{\M{J}}_k (\V{m}_{\rm t}\odot |\V{x}_i|^2)\Big]\bigg\}\nonumber\\
&=[(\V{1}-\V{m}_{\rm t})\odot \widetilde{\M{J}}_k \V{m}_{\rm t}]^{\rm T}\M{S}[(\V{1}-\V{m}_{\rm t})\odot \widetilde{\M{J}}_k \V{m}_{\rm t}]\nonumber\\
&=|a_k|^2+(\mu_4-1)a_k,
\end{align}
while the cross terms $\mathbb{E}\{(r_{l,l})^*r_{p,p}\}$ can be computed as
\begin{align}\label{cross_terms}
\mathbb{E}\{(r_{l,l})^*r_{p,p}\} &= [(\V{1}-\V{m}_{\rm t})\odot \widetilde{\M{J}}_l \V{m}_{\rm t}]^{\rm T}\widetilde{\M{J}}_l\M{S}\widetilde{\M{J}}_p^{\rm T}\
\nonumber\\
&\hspace{3mm}\cdot [(\V{1}-\V{m}_{\rm t})\odot \widetilde{\M{J}}_p \V{m}_{\rm t}]\nonumber\\
&=a_l a_p +(\mu_4-1) (a_l -b_{lp}).
\end{align}
Substituting \eqref{self_terms} and \eqref{cross_terms} into \eqref{eprgi_expanded} yields \eqref{eprgi_full}.
\end{IEEEproof}

\section{Sidelobe Analysis}
\subsection{Proof of Proposition \ref{prop:esl}}\label{sec:proof_esl}
\begin{IEEEproof}
From \eqref{range_profile}, the sidelobes can be rewritten as
\begin{align}
r_{k,l} &= [\V{0}_{N+l};\V{m}_{\rm t}\odot\V{x}_i;\V{0}_{N-l}]^{\rm H} (\M{I}_3\otimes \M{M}_{\rm r}) \nonumber\\
&\hspace{3mm}\cdot [[\V{m}_{\rm t}\odot \V{x}_{i+1}]_{N-k+1:N};\V{m}_{\rm t}\odot \V{x}_{i-1};\V{m}_{\rm t}\odot\V{x}_i;\nonumber \\
&\hspace{7mm}[\V{m}_{\rm t}\odot \V{x}_{i+1}]_{1:N-k}] \nonumber \\
&=(\V{m}_{\rm t}\odot \V{x}_i)^{\rm H}\widetilde{\M{J}}_l^{\rm H} (\M{I}-\M{M}_{\rm t}) \widetilde{\M{J}}_k(\V{m}_{\rm t}\odot \widetilde{\V{x}}_{i,k,l}),
\end{align}
where $\widetilde{\V{x}}_{i,k,l}$ is given by
$$
\widetilde{\V{x}}_{i,k,l}= \left\{
  \begin{array}{ll}
    \begin{aligned}
    &\widetilde{\M{J}}_k^{\rm H}[[\V{x}_i]_{N-k+1:N-k+l};\\
    &\hspace{3mm}[\V{x}_{i-1}]_{N-k+l+1:N};[\V{x}_i]_{1:N-k}]
    \end{aligned}, & \hbox{$k>l$;} \\
    \begin{aligned}
    &\widetilde{\M{J}}_k^{\rm H}\big[ [\V{x}_i]_{N-k+1:N};\\
    &\hspace{3mm}[\V{x}_{i-1}]_{1:l-k};[\V{x}_i]_{l-k+1:N-k}\big]
    \end{aligned}, & \hbox{$k<l$;}\\
    \V{x}_i, & \hbox{$k=l$.}
  \end{array}
\right.
$$
For data payload signals, we have
$$
r_{k,l} = \sum_{n=1}^N [\V{m}_{\rm t}\odot \V{x}_i^*]_{\overline{n-l}} [\V{m}_{\rm t}\odot \widetilde{\V{x}}_{i,k,l}]_{\overline{n-k}} (1-m_{{\rm t},n}).
$$
Thus the expected sidelobe level can be computed as
$$
\begin{aligned}
\mathbb{E}\{|r_{k,l}|^2\} &= \sum_{m=1}^{N} \sum_{n=1}^N m_{{\rm t},\overline{m-l}}m_{{\rm t},\overline{n-l}}m_{{\rm t},\overline{m-k}}m_{{\rm t},\overline{n-k}}\\
&\hspace{10mm}\times \mathbb{E}\{[\V{x}_i]^*_{\overline{m-l}}[\V{x}_i]_{\overline{n-l}}[\widetilde{\V{x}}_{i,k,l}]_{\overline{m-k}}[\widetilde{\V{x}}_{i,k,l}]^*_{\overline{n-k}}\}\\
&\hspace{10mm}\times(1-m_{{\rm t},m})(1-m_{{\rm t},n}).
\end{aligned}
$$
Note that the expectation $\mathbb{E}\{a^*bcd^*\}$, where $a=[\V{x}_i]_{\overline{m-l}}$, $b=[\V{x}_i]_{\overline{n-l}}$, $c=[\widetilde{\V{x}}_{i,k,l}]_{\overline{m-k}}$ and $d=[\widetilde{\V{x}}_{i,k,l}]_{\overline{n-k}}$, satisfies
\begin{equation}\label{fourth_moments}
\mathbb{E}\{a^*bcd^*\}=\left\{
  \begin{array}{ll}
    \mu_4, & \hbox{$a=b=c=d$;} \\
    1, & \hbox{$a=b\neq c=d$;}\\
    1, & \hbox{$a=c\neq b=d$;} \\
    0, & \hbox{$a=d\neq b=c$;} \\
    0, & \hbox{otherwise,}
  \end{array}
\right.
\end{equation}
where the fourth line follows from the rotational symmetry assumption (Assumption \ref{asu:rotsym}). Also note that when $k\neq l$ (i.e., sidelobe), the cases corresponding to the first and the third lines never happen, since $[\V{x}_i]_{\overline{m-l}}=[\widetilde{\V{x}}_{i,k,l}]_{\overline{m-k}}$ does not hold for any $m$. Therefore, we may focus on the case where the second line of \eqref{fourth_moments} holds, namely when $m=n$, which leads to the following simplified expression for the expected sidelobe level
\begin{align}\label{esl_pre}
\mathbb{E}\{|r_{k,l}|^2\} &= \sum_{n=1}^N m_{{\rm t},\overline{n-l}} m_{{\rm t},\overline{n-k}} (1-m_{{\rm t},n}) \nonumber\\
&=[\M{R}]_{k,l},~\forall k\neq l,
\end{align}
which completes the proof.
\end{IEEEproof}

\subsection{Proof of Proposition \ref{prop:eavsl}}\label{sec:proof_eavsl}
\begin{IEEEproof}
We first note that for all $k=0$ or $l=0$, $\mathbb{E}\{|r_{k,l}|^2\}=0$. This allows us to express the summation in a more compact form
$$
\sum_{k=1}^{N-1}\sum_{l>0,l\neq k} \mathbb{E}\{|r_{k,l}|^2\} = \V{1}^{\rm T}\M{R}\V{1} - {\rm Tr}\{\M{R}\}.
$$
For the first term, we have
\begin{align}\label{first_term}
\V{1}^{\rm T}\M{R}\V{1} &= N\V{1}^{\rm T}\M{F}^{\rm H}{\rm diag}(\M{F}^*\V{m}_{\rm t}^*)\M{F}(\M{I}-\M{M}_{\rm t})\nonumber\\
&\hspace{3mm}\cdot\M{F}^{\rm H}{\rm diag}(\M{F}\V{m}_{\rm t})\M{F}\V{1} \nonumber\\
&=(\V{1}^{\rm T}\V{m}_{\rm t})^2 \V{1}^{\rm T}(\M{I}-\M{M}_{\rm t})\V{1} \nonumber \\
&= \rho^2 (1-\rho) N^3.
\end{align}
The second term satisfies
\begin{align}\label{second_term}
{\rm Tr}\{\M{R}\}&=N{\rm Tr}\{\M{F}^{\rm H}{\rm diag}(\M{F}^*\V{m}_{\rm t}^*)\M{F}(\M{I}-\M{M}_{\rm t})\nonumber\\
&\hspace{3mm}\M{F}^{\rm H}{\rm diag}(\M{F}\V{m}_{\rm t})\M{F}\}\nonumber \\
&=N {\rm Tr} \{\M{F}^{\rm H}{\rm diag}(|\M{F}\V{m}_{\rm t}|^2)\M{F}(\M{I}-\M{M}_{\rm t})\}\nonumber \\
&=\rho(1-\rho)N^2.
\end{align}
Combining \eqref{first_term} with \eqref{second_term} yields \eqref{eavsl}.
\end{IEEEproof}

\subsection{Proof of Proposition \ref{prop:singer_ideal}}\label{sec:proof_singer_ideal}
\begin{IEEEproof}
It is known in the finite projective geometry literature \cite{singer1938theorem} that $\mathbb{PG}(n,q)$ has $\sum_{i=0}^n q^n = (q^{n+1}-1)(q-1)^{-1}=N$ points as well as $N$ hyperplanes. For these points and hyperplanes, one may construct an $N\times N$ matrix $\M{A}$, referred to as the incidence matrix, characterised by
$$
[\M{A}]_{i,j} = \left\{
  \begin{array}{ll}
    1, & \hbox{The $i$-th hyperplane contains the $j$-th point;} \\
    0, & \hbox{otherwise,}
  \end{array}
\right.
$$
given a specific numbering for the points and hyperplanes. Singer \cite{singer1938theorem} proved that for all $\mathbb{PG}(n,q)$, there exist at least one numbering such that the corresponding incidence matrix $\M{A}$ takes a cyclic form, i.e., in the form of \eqref{convolution_pg}. Now we may express the range response as follows
$$
\begin{aligned}
\relax[\M{R}]_{k,l} &= [\M{A}(\M{I}-\M{M}_{\rm t})\M{A}^{\rm T}]_{k,l} \\
&= ([\M{A}]_{k,:} \odot [\M{A}]_{l,:})\V{1} - ([\M{A}]_{k,:} \odot [\M{A}]_{l,:} \odot [\M{A}]_{1,:})\V{1},
\end{aligned}
$$
where $[\M{A}]_{k,:}$ is the incidence vector between the $k$-th hyperplane and all points. We note that $[\M{A}]_{k,:} \odot [\M{A}]_{l,:}$ is in fact the incidence vector between all points and the intersection of the $k$-th and the $l$-th hyperplanes. When $k=l$, the intersection is the $k$-th hyperplane itself, which contains $\sum_{i=0}^{n-1} q^n = (q^n-1)(q-1)^{-1}$ points since it is isomorphic to $\mathbb{PG}(n-1,q)$. This implies $([\M{A}]_{k,:} \odot [\M{A}]_{k,:})\V{1}=(q^n-1)(q-1)^{-1}$, which in turn yields the expression of duty cycle in \eqref{duty_cycle_singer}. When $k\neq l$, the intersection is thus isomorphic to $\mathbb{PG}(n-2,q)$, which contains $(q^{n-1}-1)(q-1)^{-1}$ points. Therefore we have
$$
\begin{aligned}
\relax[\M{R}]_{k,k} &= ([\M{A}]_{k,:} \odot [\M{A}]_{k,:})\V{1} - ([\M{A}]_{k,:} \odot [\M{A}]_{1,:})\V{1}\\
&=\frac{q^n-1 - q^{n-1} +1}{q-1}\\
&=q^{n-1},
\end{aligned}
$$
for all $k>1$, which corresponds to the mainlobes of the range response. This also suggests that these transmission masks are mainlobe-fluctuation-ideal.

Next, let us consider the sidelobes of the range response. Note that the expression $[\M{A}]_{k,:} \odot [\M{A}]_{l,:} \odot [\M{A}]_{1,:}$ can be rewritten as
$$
[\M{A}]_{k,:} \odot [\M{A}]_{l,:} \odot [\M{A}]_{1,:} = ([\M{A}]_{k,:} \odot [\M{A}]_{1,:}) \odot ([\M{A}]_{l,:} \odot [\M{A}]_{1,:}),
$$
which corresponds to the intersection of two $(n-2)$-dimensional finite projective spaces. The result can be isomorphic to either $\mathbb{PG}(n-2,q)$ or $\mathbb{PG}(n-3,q)$. In the latter case, the number of points contained in the intersection is minimal, given by
$$
([\M{A}]_{k,:} \odot [\M{A}]_{l,:} \odot [\M{A}]_{1,:})\V{1} = \frac{q^{n-2}-1}{q-1}.
$$
In light of this, the \ac{pesl} can be written as
$$
\begin{aligned}
\max_{k>0,l>0,k\neq l} [\M{R}]_{k,l} &=\frac{q^{n-1}-1-q^{n-2}+1}{q-1}\\
&= q^{n-2}.
\end{aligned}
$$
Now, substituting the duty cycle expression \eqref{duty_cycle_singer} into \eqref{lb_pesl}, we obtain
\begin{align}
\max_{k>0,l>0,l\neq k} [\M{R}]_{k,l}&\geq \left\lceil\frac{\rho(1-\rho)N^2(\rho N-1)}{(N-1)(N-2)}\right\rceil \nonumber\\
&=\left\lceil\frac{q^{2n-1}-q^n}{q^{n+1}-2q+1}\right\rceil.
\end{align}
Observe that
\begin{align}\label{bounds_pesl}
\frac{q^{2n-1}-q^n}{q^{n+1}-2q+1}&\leq \left\lceil\frac{q^{2n-1}-q^n}{q^{n+1}-2q+1}\right\rceil \nonumber\\
&\leq \frac{q^{2n-1}+q^{n+1}-q^n-2q+1}{q^{n+1}-2q+1},
\end{align}
but
$$
q^{2n-1}-q^n < q^{n-2}(q^{n+1}-2q+1) < q^{2n-1}+q^{n+1}-q^n-2q+1
$$
also holds for all $n\geq 2$ and $q\geq 2$. This implies that 
\begin{equation}
q^{n-2} = \left\lceil\frac{q^{2n-1}-q^n}{q^{n+1}-2q+1}\right\rceil
\end{equation}
holds for all $n\geq 2$ and $q\geq 2$, which in turn implies that Singer \acp{cds} achieve the lower bound of \ac{pesl} given by Corollary \ref{coro:lb_pesl}. Therefore, we may conclude that these transmission masks are simultaneously mainlobe-fluctuation-ideal and \ac{pesl}-ideal, completing the proof.
\end{IEEEproof}

\section{Properties of Slow-Time \ac{masm}}
\subsection{Proof of Proposition \ref{prop:mask_ccs}}\label{sec:proof_mask_ccs}
\begin{IEEEproof}
Let us first note that for all $k=0,\dotsc,N-1$, it follows that
\begin{align}\label{integer_k}
a_{kT} &= \sum_{n=1}^N (1-m_{{\rm t},n})m_{{\rm t},\overline{n-kT}}\nonumber\\
&=\sum_{p=0}^{L-1} \sum_{q=0}^{T-1}(1-m_{{\rm t},pT+q+1})m_{{\rm t},\overline{(p-k)T+q+1}}\nonumber\\
&=T\sum_{p=1}^{L}(1-\widetilde{m}_{{\rm t},p})\widetilde{m}_{{\rm t},\overline{p-k}}\nonumber\\
&=T\widetilde{a}_k.
\end{align}
Next, consider the difference $a_{kT+l+1}-a_{kT+l}$ for $1<l<T$, which can be written as
\begin{align}
&a_{kT+l+1}-a_{kT+l} \nonumber\\
&\hspace{3mm}= \sum_{p=0}^{L-1}\sum_{q=0}^{T-1} (m_{{\rm t},\overline{(p-k)T-(l-q)}} - m_{{\rm t},\overline{(p-k)T-(l-q-1)}}) \nonumber \\
&\hspace{20mm}\times (1-m_{{\rm t},pT+q+1})\nonumber \\
&\hspace{3mm}=T\sum_{p=1}^{L} (\widetilde{m}_{{\rm t},\overline{p-k-c_0}} - \widetilde{m}_{{\rm t},\overline{p-k-c_1}})(1-\widetilde{m}_{{\rm t},p}),
\end{align}
where $c_0=\lfloor(l-q)/T\rfloor$, and $c_1=\lfloor(l-q-1)/T\rfloor$. Observe that $c_0=c_1$ holds except for $q=l$, for which $c_0=1,~c_1=0$. This implies that
\begin{align}\label{diff_k}
a_{kT+l+1}-a_{kT+l}&= \sum_{p=1}^{L} (\widetilde{m}_{{\rm t},\overline{p-(k+1)}} - \widetilde{m}_{{\rm t},\overline{p-k}})(1-\widetilde{m}_{{\rm t},p})\nonumber \\
&=d_k.
\end{align}
Combining \eqref{integer_k} with \eqref{diff_k}, we obtain \eqref{mask_ccs}.
\end{IEEEproof}

\bibliographystyle{IEEEtran}
\bibliography{IEEEabrv,masm_ref}
\end{document}